\documentclass[]{emulateapj}

\usepackage{natbib}
\usepackage{amsmath}
\usepackage{epsfig}
\usepackage{graphicx}
\usepackage{amssymb}
\usepackage{color}
\usepackage{epstopdf}
\usepackage{rotating}
\newcommand{\e}{e$^{-}$}
\newcommand{\eps}{e$^{-}/$s}
\newcommand{\nw}{nW/m$^2$/sr}
\newcommand{\inchsign}{^{\prime\prime}}

\shorttitle{CIBER: A Sounding Rocket Payload to Study the Near IR EBL}
\shortauthors{ZEMCOV ET AL. (CIBER COLLABORATION)}

\begin{document}

\slugcomment{Submitted to ApJS September 9, 2011; accepted November 20, 2011 as part of CIBER Instrument Special Issue.}

\title{The Cosmic Infrared Background Experiment (CIBER): A Sounding
  Rocket Payload to Study the Near Infrared Extragalactic Background
  Light}

\author{
  M. Zemcov\altaffilmark{1,2}, 
  T. Arai\altaffilmark{3,4}
  J. Battle\altaffilmark{2}, 
  J. Bock\altaffilmark{2,1}, 
  A. Cooray\altaffilmark{5}, 
  V. Hristov\altaffilmark{1}, 
  B. Keating\altaffilmark{6},
  M. G. Kim\altaffilmark{7}
  D. H. Lee\altaffilmark{8}, 
  L. R. Levenson\altaffilmark{1}, 
  P. Mason\altaffilmark{1}, 
  T. Matsumoto\altaffilmark{3,7}, 
  S. Matsuura\altaffilmark{3}, 
  U. W. Nam\altaffilmark{8}, 
  T. Renbarger\altaffilmark{6}, 
  I. Sullivan\altaffilmark{9}, 
  K. Suzuki\altaffilmark{10}, 
  K. Tsumura\altaffilmark{3}, and
  T. Wada\altaffilmark{3}}

\altaffiltext{1}{Department of Physics, Mathematics and Astronomy,
  California Institute of Technology, Pasadena, CA 91125, USA} 
\altaffiltext{2}{Jet Propulsion Laboratory (JPL), National Aeronautics
  and Space Administration (NASA), Pasadena, CA 91109, USA}
\altaffiltext{3}{Department of Infrared Astrophysics, Institute of
  Space and Astronautical Science (ISAS), Japan Aerospace Exploration
  Agency (JAXA), Sagamihara, Kanagawa 252-5210, Japan} 
\altaffiltext{4}{Department of Physics, Graduate School of Science,
  The University of Tokyo, Tokyo 113-0033, Japan} 
\altaffiltext{5}{Center for Cosmology, University of California,
  Irvine, Irvine, CA 92697, USA}
\altaffiltext{6}{Department of Physics, University of California, San
  Diego, San Diego, CA 92093, USA} 
\altaffiltext{7}{Department of Physics and Astronomy, Seoul National
  University, Seoul 151-742, Korea} 
\altaffiltext{8}{Korea Astronomy and Space Science Institute (KASI),
  Daejeon 305-348, Korea}
\altaffiltext{9}{Department of Physics, The University of Washington,
  Seattle, WA 98195, USA}
\altaffiltext{10}{Instrument Development Group of Technical Center,
  Nagoya University, Nagoya, Aichi 464-8602, Japan} 
\email{zemcov@caltech.edu}


\begin{abstract}
The Cosmic Infrared Background Experiment (CIBER) is a suite of four
instruments designed to study the near infrared (IR) background light
from above the Earth's atmosphere.  The instrument package comprises
two imaging telescopes designed to characterize spatial anisotropy in
the extragalactic IR background caused by cosmological structure
during the epoch of reionization, a low resolution spectrometer to
measure the absolute spectrum of the extragalactic IR background, and
a narrow band spectrometer optimized to measure the absolute
brightness of the Zodiacal light foreground.  In this paper we describe
the design and characterization of the CIBER payload.  The detailed
mechanical, cryogenic, and electrical design of the system are
presented, including all system components common to the four
instruments.  We present the methods and equipment used to
characterize the instruments before and after flight, and give a
detailed description of CIBER's flight profile and configurations.
CIBER is designed to be recoverable and has flown twice, with
modifications to the payload having been informed by analysis of the
first flight data.  All four instruments performed to specifications
during the second flight, and the scientific data from this flight are
currently being analyzed.
\vspace{0.5cm}
\end{abstract}

\keywords{(Cosmology:) dark ages, reionization, first stars -- (Cosmology:)
  diffuse radiation -- Infrared: diffuse background --
  Instrumentation: spectrographs -- Space vehicles: instruments --
  Zodiacal dust \vspace{0.5cm}}


\section{Introduction}
\label{S:intro}

\setcounter{footnote}{0}

The extragalactic background light (EBL) is the integrated intensity
of all of the photons emitted along a line of sight through the
Universe.  Since at near infrared (IR) wavelengths the EBL is thought
to be dominated by direct emission from stars, a measurement of the
EBL at these wavelengths is an integral constraint on the total energy
released via the process of nucleosynthesis through the history of the
Universe (see \citealt{Hauser2001} for a comprehensive review).  Such
measurements are a key constraint on all models of galaxy formation
and evolution, connecting the global radiation energy density to star
formation, metal production, and gas consumption from the present back
to the epoch of reionization.

The Cosmic Infrared Background Experiment (CIBER) is a sounding-rocket
borne payload comprising four instruments designed to measure the
spectrum of the EBL, test models of the Zodiacal light intensity, and
constrain the spatial fluctuations in the EBL imprinted during the
epoch of reionization (\citealt{Bock2006}, \citealt{Zemcov2011}).  The
CIBER payload comprises two imaging telescopes (Imagers) which are
designed to study spatial fluctuations in the near IR EBL, a low
resolution spectrometer (LRS) designed to measure the
spectrophotometric properties of the EBL in the range $0.7 \leq
\lambda \leq 2.1 \, \mu$m, and a narrow band spectrometer (NBS)
designed to characterize the absolute brightness of the Zodiacal Light
(ZL) in a number of astronomical fields.  CIBER has flown twice, first
on Feb 25, 2009 and second on Jul 10, 2010, and two additional flights
of essentially the same payload configuration are planned.  A
preliminary measurement of the spectrum of the ZL as measured by CIBER
is presented in \citet{Tsumura2010}.

This paper presents the overall mechanical, cryogenic and electronic
design and implementation of the CIBER hardware.  In addition, the
customized laboratory test equipment and a description of the typical
CIBER flight profile and performance are described.  The design,
scientific motivation and implementation of the individual instruments
are presented in separate papers: \citealt{Sullivan2011} discusses the
design, implementation and performance of the Imagers, while
\citealt{Tsumura2011} describes the LRS and \citealt{Renbarger2011}
the describes NBS.  Together, these three papers are hereafter
referred to as the ``CIBER Instrument Companion'' papers.  Since the
CIBER payload was modified between its two flights to improve its
performance, throughout this paper we describe the final configuration
from the second flight, and specify the design changes informed by the
first flight.


\section{CIBER Cryogenic Design}
\label{S:mechanical}

CIBER's four instrument assemblies are seated on a common optical
bench cooled to $77 \,$K which is suspended from fore and aft
bulkheads and housed in a vacuum tight skin.  In this section we
present the mechanical and cryogenic design and implementation of the
CIBER experiment insert.

\subsection{The CIBER Payload}
\label{sS:payload}

\begin{figure*}[htb]
\centering
\epsfig{file=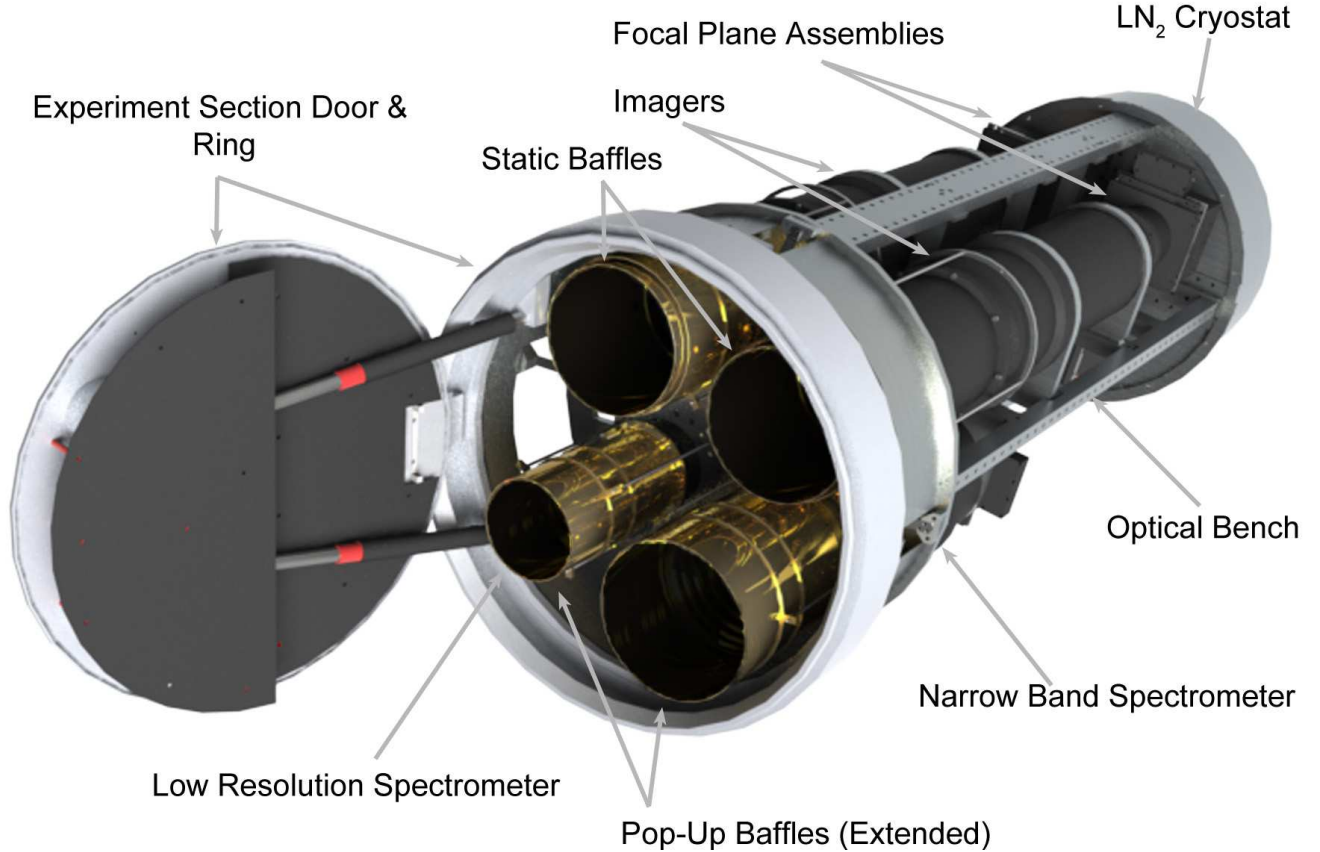,width=0.99\textwidth}
\caption{The CIBER cryogenic insert and door assembly.  The imaging
  and spectrometer instruments are mounted to an optical bench which
  is cooled by a LN$_{2}$ cryostat.  The instrument apertures, which
  are oriented towards the aft of the rocket payload, are heavily
  baffled to prevent scattered light from reaching the detectors with
  both static and `pop-up' light baffles.  When fully assembled, the
  cryogenic insert is housed in Al-1100 radiation shields which are
  connected to the cryostat and cool to $\sim 110 \,$K.  The whole
  insert is housed in an Al-6061 aluminum cylinder which, when
  connected at the fore to a pressure bulkhead and at the aft to the
  experiment door, is both vacuum tight and thermally isolated from
  the cryogenic insert section.  This vacuum cylinder also serves
  double--duty as the skin of the rocket.}
\label{fig:model}
\end{figure*}

\begin{figure}[htb]
\centering
\epsfig{file=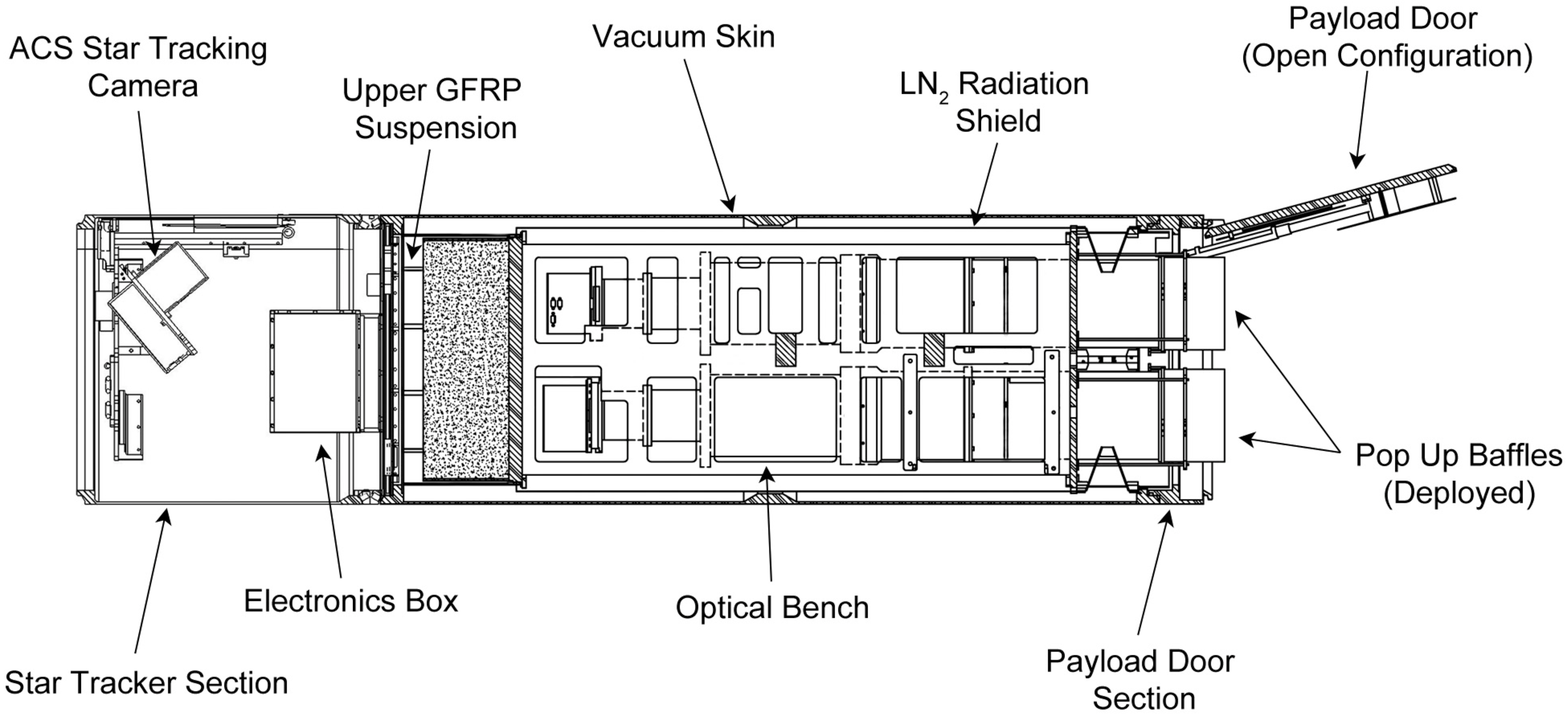,width=0.47\textwidth}
\epsfig{file=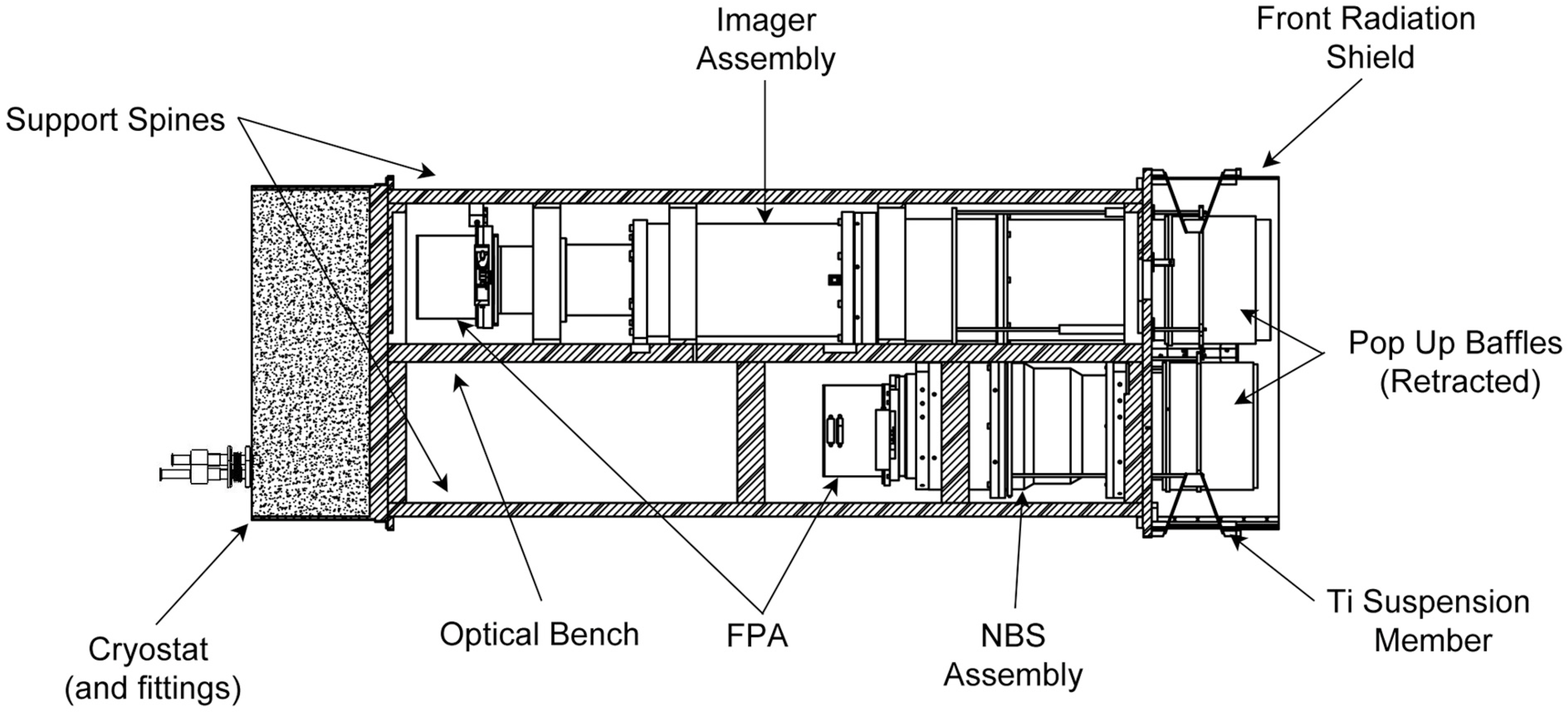,width=0.47\textwidth}
\caption{Cross sectional schematics of the CIBER payload section as
  described in the text and shown in Figure \ref{fig:model}.  Payload
  forward is to the left.  Both views show the cryogenic instrument
  insert with the telescopes installed including the cryostat, optical
  bench and suspension.  The hatched areas show sections passing
  though solid regions e.g.~the optical bench.  The upper view shows
  the experiment section door in the open configuration, the rear
  experiment bulkhead and electronics box, and the star tracker
  section housing including the ACS star tracker.  The lower view,
  which is rotated $90^{\circ}$ about the long axis to the upper,
  shows the cryogenic insert section with the upper radiation shield
  and stowed pop up baffles (see Figure \ref{fig:front_assembly} for
  detail) and the skin and radiation shield suppressed.}
\label{fig:section1}
\end{figure}

A model of the CIBER cryogenic insert including the four telescopes is
shown in Figures \ref{fig:model} and \ref{fig:section1}.  The sounding
rocket defines the coordinate system of the payload, with aft being
the side through which the instruments view, and forward in the
direction towards the experiment section's cryostat.  Except where
noted, aluminum alloy Al-6061 is used throughout the CIBER instrument
assembly.  In operation, the cryogenic insert is suspended from the
experiment section skin which acts as both vacuum jacket and
structural member of the rocket assembly (see Section \ref{S:flight}).
The experiment section is made vacuum tight by the experiment door at
the aft end, the skin, and the forward vacuum bulkhead, all of which
are manufactured by NSROC\footnote{NSROC is the NASA contractor
  responsible for the implementation and maintenance of the NASA
  Sounding Rocket Program based at NASA's Wallops Flight Facility; see
  {\tt http://sites.wff.nasa.gov/code810/nsroc.html}.}.  The skin is
gold plated on its interior to reduce its IR emissivity.  The forward
bulkhead interfaces the experiment section to the rest of the payload
and hosts the hermetic connections through which the experiment wiring
harness runs (see Section \ref{S:electronics}).

Starting at the forward end of the cryogenic insert, the experiment
section comprises a $7$ liter liquid nitrogen (LN$_{2}$) cryostat
(described in detail in Section \ref{sS:cryogenics}) with a $1300
\,$cm$^{2}$ work surface.  Just forward of the work surface are
attached twelve glass fiber reinforced plastic (GFRP) plates which
mate to the forward pressure bulkhead.  Mounted to the cryostat's work
surface is the optical bench to which the instruments are attached.
Figure \ref{fig:model}\ shows the configuration of the instruments.
All four assemblies were designed and manufactured by the Genesia
Corporation of Japan using specifications developed by the science
team.  The specifications for the instruments themselves are described
in detail in the CIBER Instrument Companion papers; here we summarize
them in Table \ref{tab:specs} and below in Sections \ref{ssS:imagers},
\ref{ssS:LRS} and \ref{ssS:NBS} for reference.
%
\begin{table*}[ht]
\centering
\caption{CIBER Instrument Specifications}
\begin{tabular}{lccc}
\hline
 & Imagers$^{a}$ & LRS & NBS  \\ \hline

Aperture & 11 cm & 5 cm & 7.5 cm  \\

Pixel size & $7" \times 7" \,$ & $1.4' \times 1.4'$ & $2' \times 2'$ \\ 

Detector array & HAWAII-1 & PICNIC & PICNIC \\ 

Pixel layout & $1024 \times 1024$ & $256 \times 256$ & $25 \times 256$ \\

FOV & $2^{\circ} \times 2^{\circ}$ & 5 slits, each $5.5^{\circ} \times 2.8'$
& $8^{\circ} \times 8^{\circ}$ \\ 

$\lambda$ & $1.0/1.6 \, \mu$m & $0.7{-}2.1 \, \mu$m & $854.2 \,$nm \\

$\lambda / \delta \lambda (=R)$ & $2$ & $15{-}25$ & $1200$ \\

Optical QE & $0.85$ & $0.8$ & $0.65$ \\ \hline

Array QE at $ 2.2 \, \mu$m & $0.7$ & $0.8$ & $0.8$ \\

Astronomical photo current (\eps) & 15 & $0.6$ & $0.5$ \\

Dark current (\eps) & $0.3$ & $<0.1$ & $<0.1$ \\

Read noise (e$^{-}$) & 15 & 26 & 28 \\

Approximate $\lambda I_{\lambda}$ Sensitivity  & & & \\

\hspace{0.5cm}  (\nw, $1 \sigma/$pix) & 30 & 50 & 90\\

\hline

Reference & \citet{Sullivan2011} & \citet{Tsumura2011} &
\citet{Renbarger2011} \\ \hline

\multicolumn{4}{l}{$^{a}$CIBER has two imagers on-board, nominally
  centered near astronomical I and H-bands.}\\
\end{tabular}
\label{tab:specs}
\end{table*}
The optical bench's surface has dimensions $83.1 \times 35.0
\,$cm$^{2}$, is machined to a thickness of $19 \,$mm, and is
light-weighted along its length.  Structural spines are mounted to the
optical bench to increase the strength of the full assembly and these
bisect the optical bench along its top and bottom to create four bays
for the instruments.  At the aft end of the optical bench is the
``front plate'' - this serves to stiffen the assembly, interface the
bench to the aperture-end suspension, and provide a stable base for
the pop up baffle assemblies.  The four instrument static light
baffles protrude through holes in the front plate which are designed
to minimize the space between the edges of the holes and the baffles.
The remaining gaps in the front plate are taped closed using
aluminized mylar tape; this makes the insert section light-tight while
minimizing heat flow from front plate to baffle.  This ``cryogenic
insert'' section is surrounded by a cylindrical radiation shield to
ensure a low radiative load on the optical bench and instruments over
most of their length (see Section \ref{sS:cryogenics}).  Fully
assembled the insert has diameter $200 \,$mm and length $590 \,$mm.

The CIBER cryogenic insert is suspended from the rocket skin by a
suspension system.  At the top end of the insert, the twelve GFRP
plates discussed above support much of the weight of the insert
section, while at the bottom the suspension centers the insert in the
skin to provide both lateral strength and $6\,$mm of vertical
compliance to take up the change in length of the assembly when it
contracts under cooling.  The insert is installed with the suspension
under compression at room temperature and provides tension when the
instruments are cooled and the optical bench has contracted.
Simulations were performed on the final optical bench and suspension
assembly design to study its harmonic and stress properties before
fabrication.  In the first flight the forward suspension was composed
of eight counter-angled GRFP plates, but these were too large to allow
installation of the pop-up baffles (see Section \ref{sS:baffling}).
In the second flight these GFRP plates were replaced with four
counter-angled Grade 5 Titanium plates with a much smaller footprint
and are visible in the payload model detail shown in Figure
\ref{fig:front_assembly}.  As a result of landing, the first flight's
upper suspension system ruptured at the bolt fixtures on both the
bulkhead and cryostat; in the second flight these interfaces were
reinforced to mitigate this problem.  In both flights, the momentum of
the cryogenic insert caused the whole assembly to bounce against the
skin at landing due to the top suspension members bending under
negative acceleration, which in turn caused the cryogenic fill tubes
to rupture.  For CIBER's third flight, bumpers have been installed on
the forward vacuum bulkhead which should constrain the motion of the
insert along the long axis of the experiment section.

\subsubsection{Wide Field Imagers}
\label{ssS:imagers}

The CIBER Imagers are optimized to measure fluctuations in the near IR
EBL arising during the epoch of reionization.  We fly two identical
Imager assemblies with different filter sets, one centered at $1.1 \,
\mu$m and the other at $1.6 \, \mu$m, chosen to bracket the peak in
the electromagnetic spectrum of the reionization fluctuations.  The
Imagers have a continuous $2^{\circ} \times 2^{\circ}$ field of view
to allow measurement of the fluctuation power at scales larger than
the expected peak in the angular power spectrum at 10 arcminutes
\citet{Cooray2004}.  The use of commercially available near IR HgCdTe
detector arrays with $1024 \times 1024$ pixels leads to a plate scale
of $7"/$pixel on the sky, which allows removal of foreground galaxies
to below the expected reionization signal level.

\subsubsection{Low Resolution Spectrometer}
\label{ssS:LRS}

The LRS is an $15 \leq R \leq 30$ spectrometer designed to measure the
near IR EBL at $0.8 \leq \lambda \leq 2.0 \, \mu$m.  The LRS comprises
an optical coupler with a slit mask at its focus coupled to an imaging
camera with a $256 \times 256$ HgCdTe detector array.  Spectral
dispersion is achieved using a prism located between the output of the
coupler and the camera.  The incident light is focused on the five
slits in the slit mask, recollimated by the coupler, dispersed by the
prism, and imaged onto the detector array.  The images of the slits
yield $5^{\circ} \times 2.8'$ cuts of the sky along one detector array
axis, and spectral dispersion along the other.  The resulting
dispersed images are cuts along the sky and, after stars are masked,
can be summed along the imaging direction and over the five slits to
produce an extremely accurate measurement of the ZL and EBL spectra.

\subsubsection{Narrow Band Spectrometer}
\label{ssS:NBS}

The final instrument in the CIBER payload is the NBS which is a fast
refractive camera with an $R=1250$ filter at its aperture.  The
geometry of the instrument and filter causes the effective wavelength
imaged at the focal plane to change with position on the detector
array \citet{Eather1969}.  By tuning the filter's bandpass to a solar
line which would be seen in reflection off the Zodiacal dust, the
absolute brightness of the ZL can be measured.  In the case of the
first and second CIBER flights the \ion{Ca}{2} at 8542\AA\ line is
chosen, though others are available.  A direct measurement of the ZL
in each CIBER field is important to the spectral measurement as it
removes the need to model the Zodiacal dust, which is a difficult
problem.  Furthermore, as the ZL is not expected to have much
structure over the NBS's $8^{\circ} \times 8^{\circ}$ field of view,
large $2' \times 2'$ pixels are sufficient to achieve the ZL
measurement.

\subsection{Liquid Nitrogen Design}
\label{sS:cryogenics}

The fundamental cryogenic element of CIBER is the $7$ liter LN$_{2}$
tank, which provides cooling for the entire cryogenic insert section.
LN$_{2}$ is selected as the cryogen for CIBER since the thermal noise
floor of HgCdTe detectors occurs near or above $80 \,$K and LN$_{2}$
is readily available and easily handled.  The calculated thermal
loading on the cryogenic insert is listed in Table \ref{tab:loading};
the interior of the cryogenic insert must be shielded against this
power to keep the components near $80 \,$K.  Failure to do so would
change the detector array properties and optical focus of the
instruments.  For this reason we employ radiation shields when testing
and in flight.  The main radiation shields cover the section of the
experiment from the cryostat, to which they are thermally connected,
to the top plate, from which they are thermally isolated.  The shields
are separated from the optical bench and spines by Vespel SP-1
standoffs at the front plate and along the optical bench spines; this
arrangement ensures that the radiative load which falls on the shields
flows directly to the cryostat.  The shields are fabricated from the
high thermal conductance alloy Al-1100 to facilitate heat flow through
the material.  These shields are wrapped in a 10 layer plastic backed
aluminized mylar blanket\footnote{NRC-2 Cryolam-C manufactured by MPI
  Technologies.} to further reduce the radiative load.  When this
blanket is installed, under typical operating conditions the inner
surface of the radiation shield operates at $\sim 110 \,$K at the
front of the experiment with a smooth gradient to $80 \,$K at the
cryostat.  The flight radiation shield is reinforced with three
cylindrical Al-6061 ribs to damp excitational modes during vibration
testing and launch, and hard black anodized on the interior surface to
act as a near-IR absorber.

\begin{table*}
\centering
\caption{Calculated cryogenic insert thermal power budget.}
\begin{tabular}{l|cc|cc}
\hline
Component & \multicolumn{2}{c|}{Optical Testing Configuration} &
\multicolumn{2}{c}{Flight Configuration} \\ \cline{2-5} 
& Emissivity & Power (W) & Emissivity & Power (W) \\ \hline
\textbf{Radiative Loads} & & & & \\ 
Top of radiation shield & 0.04  & 2.6 & 0.9 & 0.5 \\
Rear bulkhead & 0.04 & 1.6 & 0.04 & 1.6 \\ 
Radiation shields (MLI) & 0.015 & 10.5 & 0.015 & 10.5 \\
Warm optical windows$^{\mathrm{a}}$ & 1.0 & 21.5 & - & - \\
Cold optical windows & 1.0 & 5.6 & - & - \\
Shutter door & - & - & 0.9 & 3.0 \\
\hline
\textbf{Conducted Loads} & & & & \\
Upper GFRP suspension &  & 1.14 &  & 1.14 \\
Lower Ti suspension &  & 0.03 &  & 0.03 \\ 
Cryogenic service &  & 0.18 &  & 0.18 \\
Wiring harness &  & 0.16 &  & 0.16 \\
\hline
Total Power & & 22.2 W & & 17.1 W \\
Cryostat hold time & & 11.9 h & & 15.4 h \\ \hline
\multicolumn{5}{l}{$^{\mathrm{a}}$ Do not contribute to power incident
  on telescope apertures.}
\end{tabular}
\vspace{10pt}
\label{tab:loading}
\end{table*}

\begin{figure}[htb]
\centering
\epsfig{file=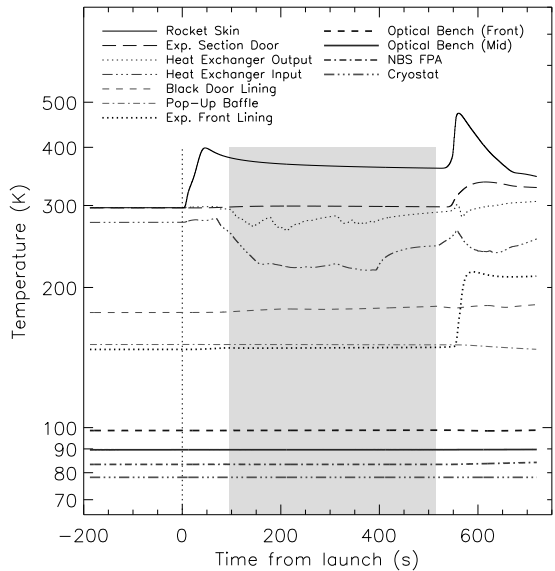,width=0.4\textwidth}
\caption{Temperatures of CIBER components during the second flight.
  Here the time axis is referred to launch at $t=0$ (highlighted by a
  dashed line).  The gray band shows the period during which
  astronomical observations occur.  The lines show the temperatures of
  various components as listed in the legend.  Thermometers near the
  outer surface of the experiment (left hand legend) dramatically
  respond to flight events like launch and reentry, while components
  protected from the ambient thermal environment (right hand legend)
  do not.  These thermometer data show that the CIBER thermal and
  cryogenic design worked well in flight.}
\label{fig:T_flight}
\end{figure}

As shown in Figure \ref{fig:T_flight}, since the main radiation
shields end at the top plate the thermal power falling on the front of
the instrument assembly causes the temperature of the optical bench to
range from $78 \,$K at the cryostat to $85{-}90 \,$K at the front
plate.  If the quiescent optical load from the skin and door sections
of $\sim 30 \,$W were allowed to fall on the instruments and optical
bench the detectors would warm up and have variable temperature which
would lead to different characteristics in testing and flight
configurations.  For this reason, the thermal shunting system
described in Section \ref{sS:opticaltesting} is used when testing the
optical properties of the instruments.

The cryostat itself must retain LN$_{2}$ under five conditions: 1)
while the experiment is horizontal during laboratory test under large
thermal loads; 2) while vertical during laboratory tests and before
launch; 3) when under $5\rm{g}-13\rm{g}$ positive acceleration during
the initial and booster rocket firings; 4) while under $-1\rm{g}$
negative gravity due to aerodynamic drag between stage firings; and
finally 5) during ten minutes of $0\rm{g}$ while observing in space.
Operation during the first three of these acceleration phases is made
possible by placing the fill and vent lines at the upper outside
corner of the cryostat.  The negative gravity period lasts $7 \,$s.
Though a small amount of liquid is lost during this period, it is
evaporated in the heat exchanger and vents from the system as a gas.
Operation in zero gravity is ensured by filling the cryostat with a
coarse aluminum foam to which the LN$_{2}$ adheres via surface tension.

\begin{figure}[htb]
\centering
\epsfig{file=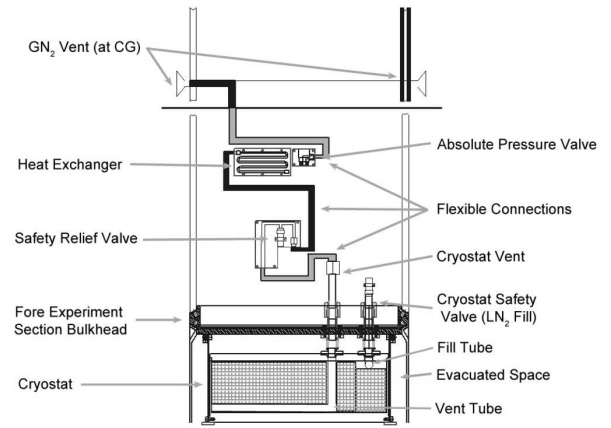,width=0.49\textwidth}
\caption{CIBER's cryogenic venting system in flight configuration.
  The vent system must operate under both positive and negative
  accelerations while ensuring the vented gas does not cause
  rotational torques to the payload section.  The $7 \, \ell$ cryostat
  contains aluminum foam to ensure good contact between the LN$_{2}$
  and experiment while in $0\rm{g}$.  The vent system has two safety
  release valves in case the experiment over-pressures on the ground
  or in flight, and an absolute pressure valve ensures the pressure in
  the cryostat is always slightly higher than one atmosphere to hold
  the cryogen at a constant temperature.  A heat exchanger ensures
  that whatever gas passes through the system is thermalized to the
  temperature of the payload before reaching the absolute pressure
  valve.  Finally, the cryogen exhaust gas is vented symmetrically
  about the center of mass of the payload to prevent unbalanced
  torques or drifts while in flight.}
\label{fig:vent}
\end{figure}

The cryogenic vent system is designed to maintain a constant cryogen
temperature both during laboratory testing operations when the external
pressure is one atmosphere and in flight when the external pressure
decreases to vacuum.  This requirement is driven both by the need for
small thermal drifts at the focal plane and to prevent air from
freezing in the vent system after re-entry.  The CIBER vent system
achieves this and is shown schematically in Figure \ref{fig:vent}.
Evaporated nitrogen from the cryostat is vented through a hermetic
connection, then through a custom heat exchanger sunk to a rocket
bulkhead to warm the vent gas to $\sim 270 \,$K, and then through a
commercial absolute pressure valve\footnote{Tavco Inc.~part number
  2391243-2-9.} which maintains a pressure of $17.5 \,$psia, i.e.~just
above one atmosphere.  The output of the constant pressure valve is
then vented through a balanced low-thrust vent situated at the center
of mass of the rocket so that the vent gas does not disturb the
attitude control system of the rocket while in flight.

The lifetime of LN$_{2}$ in the tank while ground testing is $\sim 16
\,$ h.  In practice, the cryogen tank is topped off about $2 \,$h
before launch so there are $\sim 6$ liters of LN$_{2}$ during flight.
The temperature of various components in the experiment section
immediately before launch and during the flight cycle is shown in
Figure \ref{fig:T_flight}.  There is no evidence of temperature change
at the cryostat, so we infer that the foam retention and venting
schemes work well.

\subsection{Optical Baffling}
\label{sS:baffling}

The data acquired during CIBER's first flight indicated that there was
significant thermal radiation from the hot rocket skin scattering into
the optics during astronomical observations.  This increased the
photo-current in the instruments, most prominently at the long
wavelengths of the LRS, in the $1.6 \, \mu$m Imager, and the NBS.
During ascent, the rocket skin is frictionally heated by atmospheric
drag to temperatures above $400 \,$K (see Figure \ref{fig:T_flight}),
pushing the thermal emission peak to shorter wavelengths and leading
to a large scattered signal at the detectors.  The effects of this
scattered radiation on, and the mitigation strategies for, the
individual instruments are presented in the CIBER Instrument Companion
papers; since it is common to all four instruments, the optical
baffling employed in CIBER's second flight is discussed here.

\begin{figure}[htb]
\centering
\epsfig{file=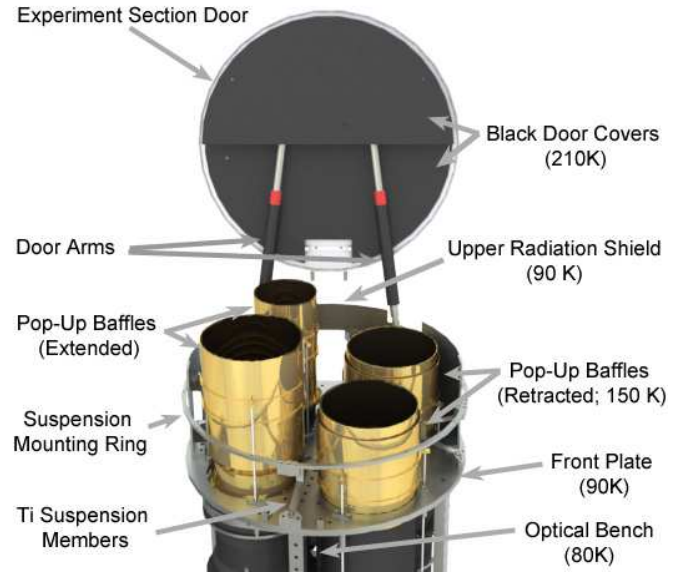,width=0.49\textwidth}
\caption{Detail view of the front of the CIBER cryogenic insert.  The
  door mounting ring and front of the upper radiation shield have been
  suppressed for clarity.  CIBER's optical baffling requirements are
  driven by the need to block all direct illumination paths from the
  instrument baffles to the door and skin.  To minimize these paths,
  pop-up baffles were installed for CIBER's second flight.  These are
  normally stowed in the retracted configuration, but are extended in
  flight by means of springs or wires to the door plate.  In addition,
  an upper radiation shield blocks most of the solid angle between the
  front plate and the edge of the door mounting ring.  Together, these
  modifications reduce the scattered thermal light contamination in
  CIBER's the second flight to below detection thresholds in all four
  instruments.}
\label{fig:front_assembly}
\end{figure}

Post-flight investigation showed that light originating at the upper
part of the skin and door can either directly, or via a single scatter
off the static baffling, illuminate the first optic of the instrument,
and from there scatter to the detector.  The simplest way to combat
this problem is to increase the height of the optical baffling and
block all light paths from the inner surface of the baffles to the
skin and door.  For CIBER's second flight, `pop-up' baffles were
installed on each of the four instruments (these are shown in detail
in Figure \ref{fig:front_assembly}).  These baffles have a slightly
larger diameter than their static counterparts and extend when the
experiment door is open so that their top edge is $\sim 0.5\inchsign$
beyond the lip of the rocket skin.  The baffles themselves are
constructed from Al 6061 and coated with Epner laser
black\footnote{See {\tt http://www.epner.com/laser\_black.ssi}.} on
their interior surface and gold plated on their outer surface.  These
baffles ride on steel rods run through brass bushings and are either
spring loaded or tied to the door with steel wire so they extend when
the door is opened.  Based on measurements of the specular and
spectral reflectivity of optically black surface coatings the Epner
laser black is $\lesssim 0.5 \,$\% reflective.  Measurements of the
instruments' sensitivity to light incident from angles $> 5^{\circ}$
have shown that the pop up baffles reduce the expected thermal pickup
by at least a factor of $10^{3}$ \citep{Tsumura2011}.

The functionality of the pop-up baffles was tested with the experiment
section in flight configuration using the vacuum box discussed in
Section \ref{sS:labsphere}.  In addition, the pop-up baffle assemblies
underwent the same vibration qualification tests as the cold shutters
as described in Section \ref{sS:FPAs} at both the component and full
assembly levels.  These tests indicate that the baffles work well
under conditions similar to the launch and flight environments.

In addition to the pop-up baffles, for CIBER's second flight
additional radiation shields were added to the front section of the
payload extending beyond the front plate to within $1 \,$cm of the
surface of the door.  These shields block $90 \,$\% of the solid angle
to the skin as viewed by the instrument static baffles and are heat
sunk to the cryostat via the optical bench, cooling to $90 \,$K.  In
addition, black caps utilized to cover the apertures of the baffles in
the door closed configuration during the first flight were augmented
to cover the entire $300 \,$K surface of the door in the second flight
assembly.  These covers are stood off from the door surface by Vespel
posts and radiatively cool to $210 \,$K, at which point their thermal
emission is undetectable.  Similarly, the pop-up baffles are thermally
separated from the front plate by the steel rods but radiatively cool
to $150 \,$K due to their close proximity to the $90 \,$K static
baffles.


\section{Focal Plane Assembly}
\label{S:focalplanes}

The CIBER detector arrays are common to all four instrument assemblies
and require electrical readout, mechanical support at the focus of
each instrument, and thermal control to ensure the dark current
remains stable during testing and in flight.  This section details the
mechanical, electrical and thermal design and implementation of the
focal plane assembly.

\subsection{Focal Plane Mechanical Design}
\label{sS:FPAs}

\begin{figure}[htb]
\centering
\epsfig{file=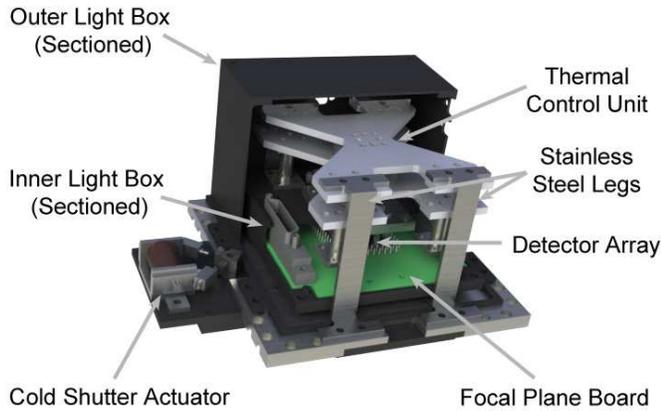,width=0.49\textwidth}
\caption{A model of a CIBER focal plane assembly (FPA) as described in
  Section \ref{sS:FPAs}.  Each of the four instruments uses an FPA to
  provide the detector array with electrical service, support at the
  focus of the instrument's optics, and thermal control.  In this
  figure the various light shields employed in the FPA have been cut
  away to show the interior structure of the unit, including the
  thermal control system and the focal plane board which provides
  electrical connections to the detector chip and preamplification
  circuit.}
\label{fig:fpa}
\end{figure}

All four instruments employ a focal plane assembly (FPA), a model of
which is shown in Figure \ref{fig:fpa}.  Beginning at the center of
the FPA, each detector array is held in commercially available chip
carrier\footnote{Part numbers AMP 641749-2 for the PICNIC arrays and
  AMP 643066-2 for the HAWAII-1 arrays.}.  This carrier is soldered
into a focal plane board (FPB; see Section \ref{sS:coldelectronics})
which electrically interfaces and mechanically supports the detector
assembly.  The FPB itself is mounted into an inner light tight box
which has a surface against which the frame of the detector chip is
pressed by a spring loaded plunger assembly.  This arrangement ensures
the chip is precisely positioned against the metal reference surface
with a calibrated force.  The spring force is selected to produce firm
pressure while also preventing mechanical stress on the detector
chips, and is set to $7.7 \,$N through a displacement of $4 \,$mm.

The inner light box is mounted to the instrument interface plate via
an assembly which provides active thermal control of the entire inner
light box.  Active thermal control is required because the on-board
detector amplifiers have a temperature dependent voltage of $\sim
1000\,$e$^{-}$/K referred to the input (R.~Smith, personal
communication).  The CIBER requirement is to stabilize detector
current drifts at the level of $0.01{-}0.1 \,$\eps\ which leads to the
specification on thermal fluctuation control of $\pm 10{-} 100 \,
\mu$K/s, depending on the instrument.  The inner light box is mounted
on an Al-1100 structure which contains a thermal choke point (see
Figure \ref{fig:fpa}).  A thermal control unit comprising an Al-1100
block with a heater element and platinum-thermometer bridge circuit
are placed at this thermal choke and provide active control of the
temperature of the inner light box (see Section \ref{sS:Tcontrol} for
a detailed presentation of this unit).

The thermal control unit is connected to the instrument interface
plate via another Al-1100 plate and four stainless steel legs which
serve to thermally isolate the inner light box assembly from the
ambient thermal environment.  An outer light box surrounds the entire
assembly, ensuring that stray IR photons cannot directly couple to the
detector array.  This assembly is mechanically mounted to the
instrument by $12 \,$mm long Vespel SP-1 posts and stainless steel
screws, and strapped directly to the LN$_{2}$ tank with custom built
Cu-10100 copper straps.  The straps comprise flexible braided wire
which is electron beam welded into copper blocks at either end.  The
blocks are clamped to the FPA and cryostat to thermally connect them.
The straps have sufficient slack to mechanically decouple the FPA from
the experiment section, which is important during e.g.~vibration
testing or launch operations.  This arrangement keeps the FPA at a
constant temperature in testing configurations which cause the optical
bench temperature to vary.
 
Each FPA also hosts a cold shutter assembly based on a flight-proven
design (see e.g.~\citealt{Wildeman1993}, \citealt{Bock1998}).  The
shutters are necessary because absolutely calibrated measurements
require precise determination of the zero-point signal in the
detector.  Cold shutters situated just in front of the CIBER detector
arrays will effectively block IR light incident on the active surface
of the detector.  The design of such a cold shutter is not trivial; it
must be operable at cryogenic temperatures, fit in a highly
constrained volume defined by the optics and detector arrays, operate
under various $g$ loads and in vacuum, and survive the shock and
vibration levels encountered in a sounding rocket flight.

\begin{figure}[htb]
\centering
\epsfig{file=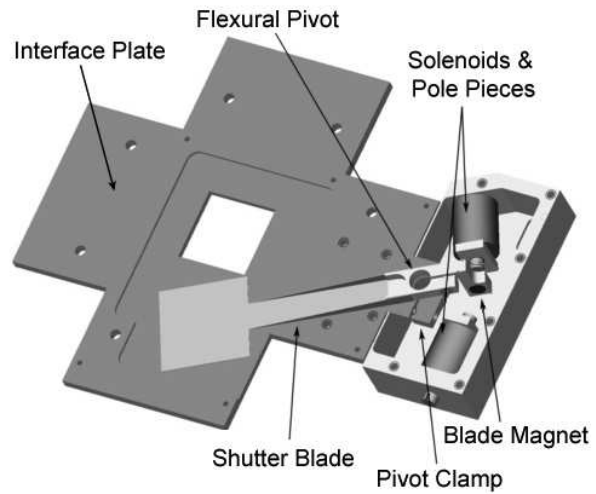,width=0.49\textwidth}
\caption{A model of a CIBER cold shutter assembly.  Each of the four
  CIBER instruments uses a cold shutter to provide a measurement of
  the absolute dark level before and during flight.  The shutter is
  actuated by changing the polarity of the current running through the
  solenoids to attract/repel the fixed magnet on the shutter.}
\label{fig:shutters}
\end{figure}

Figure \ref{fig:shutters} shows a model of the CIBER flight shutters.
The CIBER shutters are composed of a $0.005 \inchsign$ thick hard
black anodized Al-1145 blade mounted on a $0.25 \inchsign$ single
ended flexural pivot and counter weighted with a small rare-earth
magnet.  The total weight of the blade assembly is $2.4 \,$g of which
$0.4 \,$g is due to the fixed magnet and $0.1 \,$g to the blade.
Flexural pivots\footnote{Flexural pivots are precision rotational
  spring bearings actuated by flexing blades held inside a rotating
  barrel; the parts used in CIBER are manufactured by Riverhawk
  Flexural Pivots Co.~part number $5008{-}800$.} are chosen as they
are frictionless and so well-suited to cryogenic operation.  The
flexural pivot is mounted to a base plate which interfaces the shutter
assembly to the focal plane assembly.  The shutter is actuated using a
permanent magnet attached to the rear of the shutter and an
electromagnetic yoke mounted to the base plate.  The yoke comprises
two $1.5 \times 10^{4}$ turn solenoids through which a high magnetic
permeability mu-metal pole pieces are run, configured with one
solenoid on either side of the shutter throw.  When current passes
through the solenoids, the pole pieces attract or repel the fixed
magnet on the shutter in coordination, thereby actuating it.  The
magnets and yoke are configured to make the shutter bi-stable so that
the blade remains either open or closed in the unpowered state.

To test the flight-worthiness of the assembly, the cold shutters flown
on CIBER have undergone a rigorous component-level test regime
including cryogenic life cycle testing and vibration qualification.
The cryogenic lifetime test requires 5,000 open/close cycles on a test
assembly held at $77 \,$K; no change in the performance of the shutter
is observed.  The vibration test sequence follows the standard NASA
vehicle level 2 specification (see the NASA Sounding Rocket Handbook
\nocite{SRPH2005}) and involve random and sine vibration tests in the
$x,y$ and $z$ axes.  The tests have $10 \,$s duration with an
amplitude of $12.7 \, \rm{g}_{\mathrm{rms}}$.  During these tests the
shutter is clamped in the open position by powering the electromagnet,
which simulates launch conditions.  These tests are important as the
first CIBER cryogenic shutter design, which utilized a double-ended
$1/8\inchsign$ flexural pivot\footnote{Manufactured by Riverhawk
  Flexural Pivots Co.~part number $6004{-}800$.}, failed during
vibration testing due to cracking of the vanes inside the pivot
assembly.  This cracking occurred during testing at modest vibration
levels and was not anticipated based on engineering calculations,
showing that the fatigue and maximum tolerable stress margins needed
to be larger than a static stress calculation suggests.  In the second
(flight) design the shutter bearing mechanism was redesigned to use a
single ended flexural pivot with a $1/4\inchsign$ diameter as
described above; this design was successfully qualified at component
and payload level and have flown on both CIBER flights.  Before and
after each test the shutter assembly is actuated through an open/close
cycle 5 times, and during each of the tests the test assembly is
cooled to $\sim 100 \,$K and clamped in the open position by applying
the appropriate current to the electromagnet.  Finally, one of the
second-generation shutter assemblies was tested to failure; this test
showed that the flight shutters have a virtually indefinite life span
in the acceleration environment encountered in flight and only failed
at $\rm{g}$ loads twice the launch vibration specification while
unclamped and for a test duration $30$ times the requirement.  Most
importantly, the same four cold shutters have been flown twice in
CIBER with no modification or repair required following flight,
verifying that the design and implementation is sound.

\subsection{Detectors and Cryogenic Electronic Systems}
\label{sS:coldelectronics}

CIBER utilizes four Mercury-Cadmium-Telluride (HgCdTe) detector arrays
in total; two engineering-grade HAWAII-1 detector arrays for the
Imagers and two MBE-grown substrate removed PICNIC arrays for the
Spectrometers (see \citealt{Beletic2008} for details).  The basic
detector characteristics are given in Table \ref{tab:specs}.  These
detector types, fabricated by Teledyne Scientific \&
Imaging\footnote{See \tt{http://www.teledyne-si.com/}.}, have a proven
track record of low dark current, high quantum efficiency (QE), good
uniformity across the array, and low noise.  Both array types have
similar intrinsic wavelength bandpasses with $>50 \%$ QE for $0.8
\lesssim \lambda \lesssim 2.5 \, \mu$m.  CIBER's HAWAII-1 and PICNIC
arrays have $18 \, \mu$m and $40 \, \mu$m pixel pitches, respectively.
The detector material is indium bump-bonded to the source-follower
configured readout integrated circuit (multiplexer) which leads to
typical full well depths of $10^{5}$\e\ and readout noise $< 15 \,$\e.
The CIBER Instrument Companion papers detail the design requirements
leading to the choice of detector array type for each CIBER
instrument.

\begin{figure}[htb]
\centering
\epsfig{file=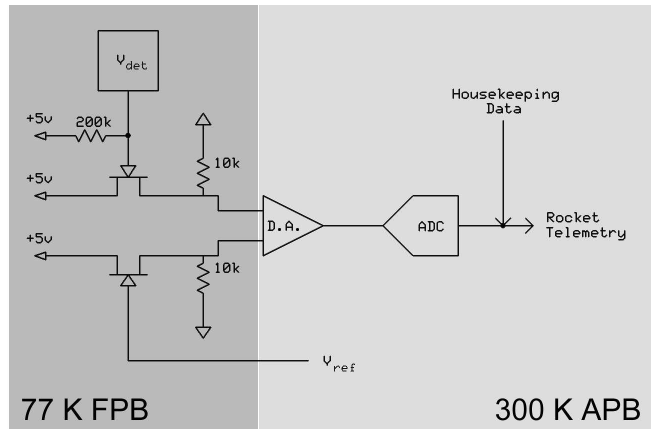,width=0.49\textwidth}
\caption{A schematic of CIBER's detector readout chain.  The $77 \,$K
  focal plane board (FPB) outlined in dark gray is designed to provide
  preamplification for both the detector voltage $V_{\mathrm{det}}$
  and the reference voltage $V_{\mathrm{ref}}$ according to the scheme
  described in \citet{Hodapp1996}; see Section
  \ref{sS:coldelectronics} for details.  The $300 \,$K array processor
  board (APB) uses a differential amplifier to difference
  $V_{\mathrm{ref}}$ and $V_{\mathrm{det}}$, which effectively removes
  zero point offsets and common mode drifts.  This differential signal
  is then converted to a digital level using an analog to digital
  converter (ADC) and housekeeping data is incorporated into the time
  stream as described in Section \ref{sS:warmelectronics}.  Finally,
  this data stream is passed to the rocket telemetry system and is
  telemetered to the ground station in real time as described in
  Section \ref{sS:TMGSE}.}
\label{fig:readout}
\end{figure}

Figure \ref{fig:readout} is a schematic representation of the CIBER
detector readout chain, which is divided into cryogenic and ambient
temperature sections.  In order to reduce electroluminescence from the
output amplifiers on the multiplexer, the output bus of the detector
array is connected directly to an external preamplifier stage
following the scheme of \citet{Hodapp1996}; this replaces three MOSFET
devices in the multiplexer with a single external JFET, which reduces
the self-emission from the multiplexer.  The CIBER readout scheme also
employs a differential amplification step; a reference signal is
generated in the warm electronics (see Section
\ref{sS:warmelectronics}) and sent to the FPB, where it is connected
to one of the inputs of a dual input JFET\footnote{CIBER uses a Vishay
  SST406-E3 dual N-channel JFET switch.}; the output bus from the
array is connected to the other.  This scheme provides common-mode
rejection to noise and pickup on the wiring between the FPA and warm
electronics, and reduces any thermal dependence on the JFET.

Both the HAWAII-1 and PICNIC detectors have four quadrants and four
independent output buses, so a FPB contains four of the unit cell
preamplification circuits shown in Figure \ref{fig:readout} to service
each detector.  In addition, the FPB provides the electrical interface
to the various clock signals required by the multiplexer for the warm
electronics where the clock signals are generated.  Each quadrant in
the HAWAII-1 arrays acts as an independent unit, meaning that the
clock signals must be generated for each quadrant, while in the PICNIC
arrays the clocking is common to all four quadrants.  A second, and
more important, difference between the HAWAII-1 and PICNIC FPBs is the
detector substrate in a PICNIC array is not held to ground so its
voltage must be optimized to minimize the dark current.
\citet{Lee2010} give a detailed account of the bias settings of
CIBER's PICNIC arrays.

\subsection{FPA Thermal Control}
\label{sS:Tcontrol}

As discussed in Section \ref{sS:FPAs}, the temperature of the focal
plane assembly and detector is actively controlled using a two-stage
thermal guard design.  The NBS has the most stringent dark current
stability requirement of $0.01 \,$\eps, which leads to a temperature
fluctuation stability requirement of $\pm 10 \, \mu$K/s.  To meet this
requirement, the CIBER thermal control unit consisting of a cryogenic
thermometer/heater pair connected to a room temperature readout and
proportional-integral-derivative (PID) controller is used.  The
cryogenic thermometer/heater pair is contained on a unit mounted to an
$81 \times 68 \times 3 \,$mm$^{3}$ Al-1100 block.  We use a Platinum
resistive thermometer\footnote{Manufactured by Sensor Scientific part
  number P200BMD is used.} with resistance $2000 \, \Omega$ at $0 \,$C
coupled to a preamplifier\footnote{Analog devices part number AD8630.}
with $30 \,$nV/$\sqrt{\mathrm{Hz}}$ input noise on minute time scales
to measure the temperature of the block.  The thermometer is read in a
Wheatstone bridge designed to balance at $80 \,$K; the bridge is
biased by a $2 \,$V low noise reference, which leads to a thermal
sensitivity at balance of $\sim 7 \,$mV/K.  The heater is a standard
thin film resistor with a value of $500 \, \Omega$.

\begin{figure}[htb]
\centering
\epsfig{file=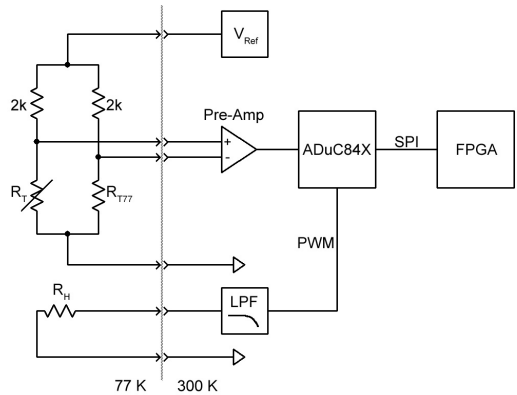,width=0.4\textwidth}
\caption{Schematic representation of the temperature control circuit
  used to actively control the thermal environment of CIBER's
  detectors.  This system consists of a voltage-biased bridge circuit
  which includes the Pt-resistance thermometer $R_{\mathrm{T}}$ and a
  differential preamplifier.  The resistance $R_{\mathrm{T77}}$ is
  selected to balance the bridge $\sim 80 \,$K, corresponding to a
  value of $\sim 350 \, \Omega$.  This signal is then digitized and
  both sent to the housekeeping board FPGA for incorporation into the
  data time stream and used to calculate PID loop parameters which are
  input to the heater control circuit.  This circuit consists of a
  variable voltage source which is low pass filtered and sent to the
  FPA to actively heat the detector unit. }
\label{fig:tc_circuit}
\end{figure} 

The bridge readout and PID control are handled by a single
device \footnote{Analog Devices ADuC84X series controllers are used.}
chosen for its 24-bit ADC and 16-bit DAC resolution, chop-stabilized
gain stage, and compact, integrated functionality.  The micro
controller is programmed to implement the Takahashi-type PID; the
parameter settings are optimized using the Zeigler-Nichols criteria by
determining the critical gain and oscillation frequency of the
proportional control loop.  Each CIBER FPA uses two such temperature
control units, one which monitors the focal plane temperature and one
which regulates the temperature control point (TCP; see Section
\ref{sS:FPAs}).  As running both units in control mode could induce
feedback effects in the temperature of the FPA unit, only the TCP unit
is run in PID-mode; the focal plane unit is programmed to pre-heat the
light box to within $0.5 \,$K of the set point and simply monitor its
temperature stability following that.  Since the temperature set point
must be larger than the highest normal bath temperature, the set point
of each FPA is optimized independently; all lie between $81$ and $83
\,$K.  The PID loop parameter tuning and pre-heating cycle were
optimized and tested in the lab prior to instrument deployment.

\begin{figure}[htb]
\centering
\epsfig{file=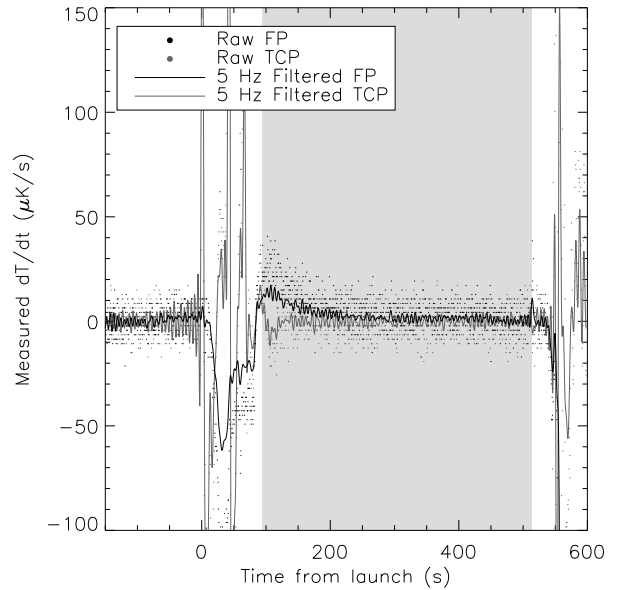,width=0.4\textwidth}
\caption{Temperature gradient of the NBS focal plane measured prior
  to, and during, CIBER's second flight.  The focal plane temperature
  is shown both as a raw data stream (black points) and filtered with
  a $5 \,$Hz low pass 1-pole Butterworth filter (black line); the same
  are also shown for the thermal control point temperatures (gray
  points and line).  Before launch the thermal stability of the FPA is
  excellent with $\sim 1 \, \mu$K/s RMS.  Launch causes large
  disturbances in the temperature control circuit, but these are
  effectively controlled by the beginning of astrophysical
  observations (shaded gray band).}
\label{fig:fptc}
\end{figure}

On flight day, the focal plane temperature control system is powered
on several hours before flight to ensure that each focal plane is
controlling at its set point at launch.  Due to its mass, the heating
time of the FPA is quite long; typically $1.5{-}2 \,$h are required
for the heaters to raise the temperature of the focal plane by the
required $\sim 2 \,$K.  Figure \ref{fig:fptc} shows the derivative of
both focal plane thermometers for the NBS instrument just prior to
launch and during flight.  These data show that, though large thermal
impulses are visible, the thermal drift during flight is $\lesssim 10
\, \mu$K/s, corresponding to a dark current drift of $\lesssim 0.01
\,$\eps, which meets our requirements.

\subsection{Calibration Lamp Assembly}
\label{sS:callamps}

\begin{figure}[htb]
\centering
\epsfig{file=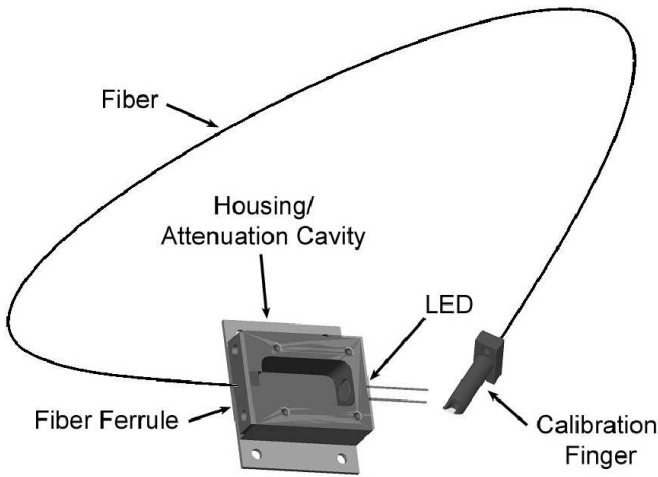,width=0.49\textwidth}
\caption{A schematic representation of a single CIBER calibration lamp
  assembly.  An LED is fiber-coupled to an instrument by means of a
  fiber and calibration finger which projects into the optical
  assembly, illuminating the detector while remaining out of the image
  of the sky.  As LEDs at normal operating powers generate photo
  current levels significantly brighter than the target of $\sim 100
  \,$\eps, the output of the LED is attenuated by the geometry of the
  coupling housing.}
\label{fig:callamp}
\end{figure}

Calibration lamp assemblies are used to monitor the response of the
detectors to a known light input before and during flight.  Each of
the four assemblies, an example of which is shown schematically in
Figure \ref{fig:callamp}, consists of an LED coupled to an optical
fiber which routes the light to an instrument.  The LEDs are selected
to have operating temperature-adjusted wavelength outputs which peak
close to the operating wavelengths of the instruments, namely $1.1 \,
\mu$m for the short wavelength Imager and NBS and $1.45 \, \mu$m for
the long wavelength Imager and LRS\footnote{These lamps are part
  numbers L7866 and L8245 from Hamamatsu Photonics, respectively; our
  testing verifies that the light output of these devices is stable
  over hundreds of cryogenic cycles.}.  For normal lamp power outputs
of $> 5\,$mW the lamps are as much as a factor of $10^{4}$ too bright
to be directly coupled to the instruments at the goal photo-current of
$\sim 1000 \,$\eps\ per detector pixel, which would provide a signal to
noise ratio of 100 in $10 \,$s of integration.  Geometric attenuation
is established at the lamp housing and is tuned for each instrument
using the distance between the LED and the fiber ferrule.  The housing
itself is attached to the optical bench spine, while the output side
of the fiber is coupled to the instrument's optics by means of a small
finger with a mirror mounted at its exit aperture.  Light emitted by
the LED is fed through the fiber and finger and illuminates the
detector array in a non-uniform, but repeatable, way.

\section{Warm Electronics}
\label{S:electronics}

CIBER's electronic system is responsible for three main functions: 1)
it biases and reads out the detector arrays; 2) it digitizes the
analog detector outputs, incorporates the housekeeping data, and
passes the resulting data stream to the rocket telemetry system; and
3) it receives the telemetered data and stores it to disk.
Additionally, the electronics drive functionality for thermal control,
cold shutters and calibration lamps and incorporate the status of
payload systems into the housekeeping data.

\subsection{Warm Payload Electronics}
\label{sS:warmelectronics}

CIBER's warm electronic system has several functions which can be
broken down into three main categories: 1) clocking and reading out
the detector array; 2) actuating systems inside the experiment
section, and 3) monitoring and reporting the status of the payload.
These are accomplished using a combination of two types of boards,
array processor boards (APB) and housekeeping boards (HKB).  Each
instrument requires two APBs and one HKB to operate, so that the
payload requires four HKBs and eight APBs in total.  These boards are
all housed in a single electronics box which is attached to the
non-evacuated side of the forward vacuum bulkhead.  When the payload
is fully assembled, this means that the electronics box protrudes into
the star tracker section (see Section \ref{S:flight} for details),
which imposes stringent constraints on the size of the electronics
box.  The overall dimensions of the electronics box are $25.2 \times
18.5 \times 17.0 \,$cm; the electronics boards are restricted to a
footprint of $18.0 \times 12.8 \,$cm with a maximum height of $2.0
\,$cm as a result.

Instrument power from the rocket batteries is supplied at a nominal
rating of $\pm 14.4$V.  Since a variety of voltages are required by
various subsystems in the CIBER electronics, the input power must be
regulated on both the APBs and HKBS.  An additional complication is
that the battery power varies from $\sim 14.5$V when the payload
switches to internal power at T$=-3$ minutes to $\sim 11.0$V at the
T$=+12$ minutes experiment power off command.  We have verified that
the CIBER electronics system operates normally for $V > 8$V.

The CIBER boards are interfaced to the cryogenic electronics by
customized cabling.  The warm cabling\footnote{Manufactured by
  Stonewall Cables.} comprises shielded copper wire connecting
standard D-subminiature connectors on the board side and 66-way
circular connectors on the hermetic side.  The cryogenic
cabling\footnote{Manufactured by Tekdata Interconnect Systems.}
comprises twisted pair maganin wire woven into a Nomex braid
connecting a 66 way circular hermetic connector and D-microminiature
connectors at the FPA and housekeeping service connections.  The
cables harnesses are $\sim 1.5 \,$m long measured from the electronics
box to FPB.

\subsubsection{Array Processor Boards}
\label{ssS:APB}

The APBs have three functions: 1) generation of detector clocking
signals synchronized with the serial clocks sent from the rocket
telemetry (TM) system; 2) generation of detector bias voltages; and 3)
amplification and digital conversion of the signals from the detector
array.  The first function is handled by a complex programmable logic
device\footnote{Lattice part number M4A5-192/96 is used.} (CPLD) which
clocks the detector array based on synchronization clock pulses from
the rocket TM.  For the Imagers, the two APBs which service the
detector create clocks independently from the same synchronization
pulses, while for the Spectrometers only a single board handles the
array clocking.

The generation of the array bias signals is handled by voltage
followers\footnote{AD8629 operational amplifiers coupled to the
  $+5\,$V power supply are used in CIBER.}; these have variable levels
which are set to values which minimize the amplification noise while
maximizing the dynamic range by setting the value of fixed resistors.
The specific choices of these bias signal voltages can be found in
\citet{Lee2010} and the CIBER Companion Instrument papers.

The detector array output conditioning is shown schematically in
Figure \ref{fig:readout}.  The amplification is handled by a
instrumentation amplifier circuit with a gain of 10.5, a bandwidth of
$5 \,$MHz and a settling time of $400 \,$ns.  Both the array output
and voltage reference are amplified in the same way with equivalent
circuits.  These signals are input to a 16-bit, $500,000$
sample-per-second differential analog to digital (A/D)
converter\footnote{Analog Devices part number AD7688 is used.} which
converts the difference of the array signal and reference voltages to
digital levels.  The digitized signals are then formatted into the
output data stream by the APB CPLD and passed on to the rocket TM
system for transmission.

The data transmission rate, and thus the clock rate of the detector,
is set by the rocket TM system's 10 Mbps serial data transfer rate.
This maximum data transmission rate leads to a single clock cycle
period of $100 \,$ns; each pixel is 16 bits so the pixel visit time is
$1.6 \, \mu$s.  The main synchronization pulses occur at $9191 \,$kHz
for the Imagers and $7812 \,$kHz for the Spectrometers.  In each
interval, either sixteen pixels for each of the four quadrants from a
single instrument (for the Imagers) or eight pixels from each of the
four quadrants from both instruments (for the Spectrometers) are
addressed.  Table \ref{tab:readout}\ accounts for the various readout
clock cycles to lead to a master read rate of $1.786$ s/frame for the
Imagers and $0.264$ s/frame for the Spectrometers.

\begin{table*}[htb]
\centering
\caption{Frame rates for the Imagers and Spectrometers.}
\begin{tabular}{l|cl|cl}
\hline & \multicolumn{2}{c|}{Imagers} & \multicolumn{2}{c}{Spectrometers} \\ 
& Data Size & Read Time & Data Size & Read Time \\ \hline 
1 Pixel & 16 Bits & $1.6 \, \mu$s & 16 Bits & $1.6 \, \mu$s \\ 
1 Packet & $64 (+4)^{\mathrm{a}}$ Pixels & $108.8 \, \mu$s & $64 (+16)^{\mathrm{a}}$ Pixels & $128 \, \mu$s \\ 
1 Line & 32 Packets & $3482 \, \mu$s & 16 Packets & $2048 \, \mu$s \\ 
1 Frame & $512 (+1)^{\mathrm{a}}$ Lines & $1.786 \,$s & $128 (+1)^{\mathrm{a}}$ Lines & $0.264 \,$s \\ \hline 
\multicolumn{5}{l}{$^{\mathrm{a}}$Numbers appearing in brackets denote
  equivalent data breadths for}\\
\multicolumn{5}{l}{rocket and experiment housekeeping data; these are 
  interleaved} \\
\multicolumn{5}{l}{into the transmitted time streams.} \\ 
\end{tabular}
\label{tab:readout}
\end{table*}

\subsubsection{Housekeeping Boards}
\label{ssS:HKB}

In addition to a master/slave APB pair, each instrument is serviced by
a Housekeeping board (HKB).  Again, the hardware design of these
boards is common to all the instruments, though the firmware
programming is different for the Imagers and Spectrometers.  The CIBER
HKB has five functions: 1) detector array reset signal and
housekeeping data generation; 2) cold shutter control; 3) calibration
lamp control; 4) temperature monitoring of components in the
experiment section; and 5) temperature monitoring and control for the
FPAs.  The last of these is discussed in Section \ref{sS:Tcontrol}; we
present the other four functions below.

The heart of the CIBER HKB is a mid-capacity field programmable gate
array\footnote{Lattice part number LCMXO2280.} (FPGA).  This device
receives the synchronization pulses from the rocket TM and is
programmed to respond to asynchronous external events by driving
hardware and sending asynchronous reset pulses to the APBs.  Array
reset commands are asserted to the APBs on open or close shutter
commands, calibration lamp enable or disable commands, or in response
to the attitude control system (ACS) computer inputs.  The ACS system
(discussed in detail in Section \ref{S:flight}) has two lines to
communicate with the CIBER HKBs corresponding to `close to target' and
`on target'.  The `close to target' status is triggered when the ACS
computer determines that the errors are less than 100 arcsec in
$(\alpha,\delta)$, for 1 second.  The `on target' status is triggered
3 seconds after `close to target' is asserted; this is the typical
time required for the ACS to settle to within 1 arcsec of the target
position.  The arrays are not reset while `on target' is asserted
(except in response to hardware interrupts like a shutter close
command; see Section \ref{S:flight}).

In addition, the experiment section door status is sent to the HKBs;
the FPGA logic is programmed to respond to this signal in the
following way.  On the ground (in the laboratory, pre-launch, or on
ascent), resetting based on the ACS `close to target' and `on target'
signals is disabled.  The experiment `door open' signal is received
roughly $80\,$s into the flight when the payload has separated from
the motor stages and is ready for observation.  Upon receiving the
`door open' status, the HKBs enter a state where the detector arrays
are reset every other frame \emph{unless} the `on target' status is
asserted.  This scheme ensures that during slews the array is reset
frequently to erase bright star trails falling across the image.
During integration no resets are asserted.  As a fail safe, the APBs
are hard coded to reset every $50 \,$s if a reset command has not been
asserted more recently, which provides useful data acquisition in the
situation that communication with the HKB is broken.

Additionally, the CIBER HKBs generate housekeeping status data which is
inserted in the 513$^{\mathrm{th}}$ (Imagers) or 129$^{\mathrm{th}}$
(Spectrometers) data line as shown in Table \ref{tab:readout}.  These
data include the status of the cold shutters, experiment door, ACS
signals, rocket event timer, reset status, thermometry and time
information.  The housekeeping data are parsed out of the data stream
on the ground and provide checks of the status of the experiment for
each frame.

The shutter and calibration lamp drivers are controlled by external
events via the FPGA, providing asynchronous control of the state of
the cold shutters and on-board calibration lamps (see Section
\ref{sS:payload}).  The calibration lamp driver is simply a constant
current source, the value of which is $\sim 1 \, \mu$mA but varies by
a factor of two between instruments.  This current source is designed
to be stable and has peak to peak variation $ < 1:10^{4}$ per hour.
The shutter driver circuit provides active differential $+\-12$V to
the shutter electromagnet such that in quiescent operation these
cancel at the solenoid.  When the `shutter open' or `shutter close'
command is received, the corresponding voltage is dropped to zero and
the shutter actuates.  The calibration lamp power and shutter commands
are monitored and transmitted in the housekeeping data for each frame.

The CIBER HKBs provide six channels of thermometry each, for a total of
24 diode devices on board.  The most important of these are plotted in
Figure \ref{fig:T_flight} for CIBER's second flight.  The diodes are
biased at a constant $10 \mu$A with a constant current source and the
voltage drop over them is measured to determine the temperature.  The
A/D conversion is performed by a controller which can handle all six
devices simultaneously\footnote{Analog Devices ADuC84X series.}.  This
chip also handles the asynchronous communication with the HKB FPGA,
which then inserts the thermometry data into the housekeeping data
stream.

\subsection{Telemetry \& Ground Station Electronics}
\label{sS:TMGSE}

The telemetry (TM) system used on CIBER is custom-built from
components designed and fabricated in-house at NSROC.  The TM system
takes the data generated by the CIBER warm electronics as input,
populates the extra 4 (Imagers) or 16 (Spectrometers) words identified
in the second row of Table \ref{tab:readout} with rocket payload
status data, and transmits it to the ground station.  The CIBER
payload's TM section utilizes three $10 \,$Mbps down links, one for
each Imager and one for the combination of both Spectrometers.  The
encoded data streams are transmitted over wrap-around antennas fitted
in recesses on the rocket skin.  The NASA ground station decodes the
data stream and stores it to tape; the experiment data can
additionally be passed to the experimenter real time.

\begin{figure*}[htb]
\begin{center}
\epsfig{file=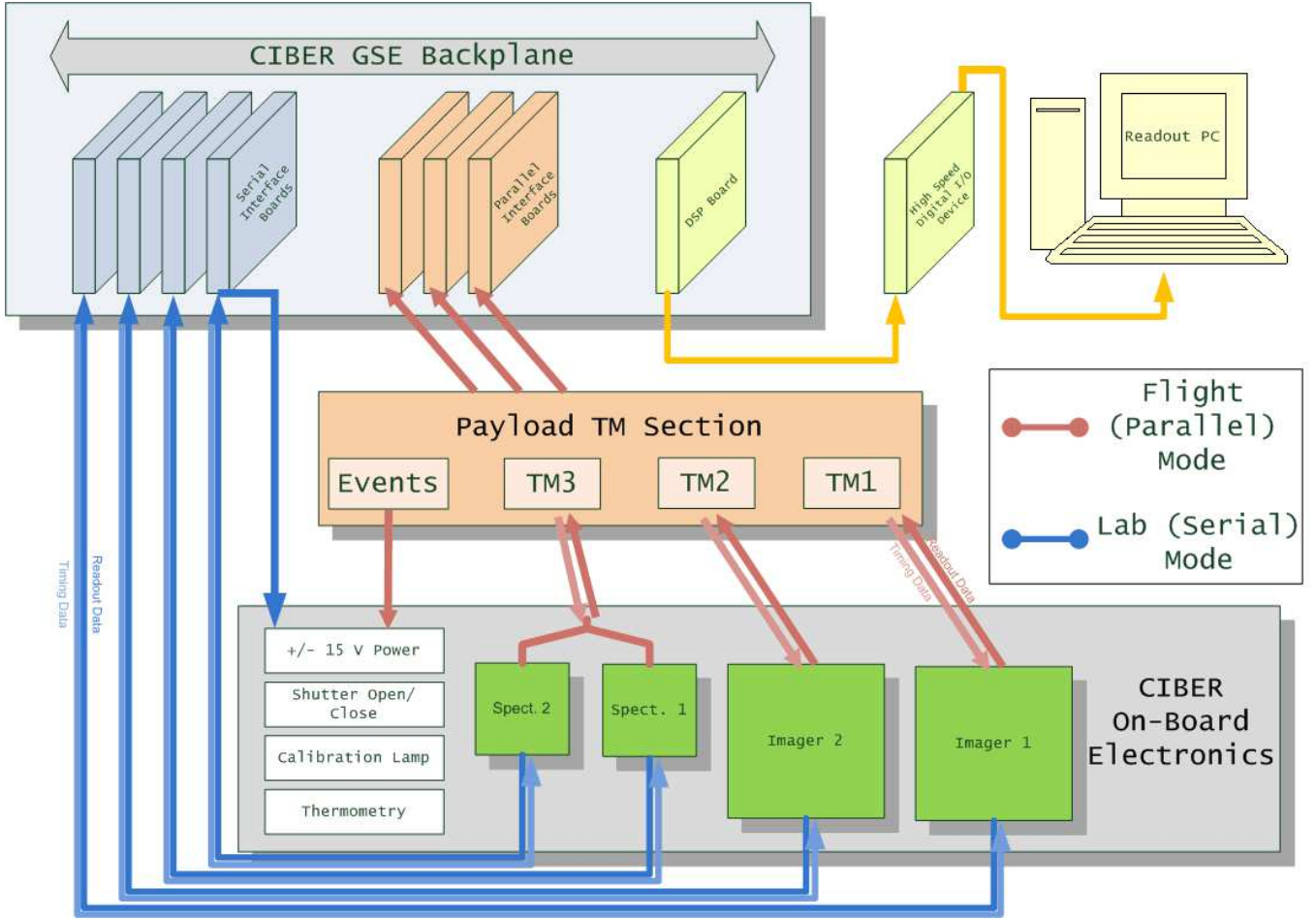,width=1.0\textwidth}
\caption{A schematic block diagram of the CIBER Ground Station
  Electronics (GSE) system.  The system has been broken into the
  on-board electronics, the payload TM system, and the CIBER GSE.  The
  red-colored lines and boxes show the data flow for the parallel
  communication mode used in flight, while the blue show the data flow
  for the serial communication used in the laboratory.  The serial mode
  is designed to be a drop-in replacement for the TM section and
  provides all of the functionality of the experiment TM so that the
  instruments can be tested without the TM section present.  The
  blocks shown in the green represent the CIBER on-board electronics
  system which include two $1024 \times 1024$ pixel Imagers and two
  $256 \times 256$ pixel Spectrometers.  In parallel mode three $10
  \,$Mbps links (TM1, 2, 3) are required to telemeter the data streams
  to three interface cards in the GSE which are used to acquire the
  data.  In serial mode one serial interface card is used per
  instrument to simulate the interface with the rocket TM.  In serial
  mode, the power and event status signals are also provided to the
  CIBER warm electronics.  In either mode, the DAQ cards communicate
  with a digital signal processing (DSP) board via a VMEbus back
  plane.  The DSP board is connected to a computer via a high speed
  digital I/O PCI device.  This system provides the capability to
  store and display real time streaming data from either the rocket TM
  or directly from the CIBER warm electronics, as well as the ability
  to simulate the functions of the TM system's timer and status
  communication.}
\label{fig:gse}
\end{center}
\end{figure*}

CIBER requires its own ground station electronics (GSE) system to
parse, display and store the data.  Importantly, this system does
double-duty as a data acquisition system for ground testing when the
experiment is not connected to the rocket TM.  The GSE system is
graphically summarized in Figure \ref{fig:gse}.  It comprises
customized data acquisition (DAQ) cards featuring mid-capacity FPGAs
which parse the incoming data; these communicate to digital signal
processing (DSP) board over a VMEbus back plane.  The front bus of the
DSP board is in turn connected to a commercial high speed digital I/O
PCI card\footnote{National Instruments part number PCI-6534.}
installed in a readout PC.  The PC runs customized software to store
the data stream to files and display it in a graphical interface in
real time.

In order to facilitate both of its roles, the GSE operates in two
modes called `Parallel Mode' and `Serial Mode' which are shown
schematically in Figure \ref{fig:gse}.  These correspond to a passive
mode where the data stream input from the NASA ground station is both
displayed real time and stored to disk, and an active mode which
simulates the functionality of the payload TM section in addition to
its normal display and data storage functionality.  In serial mode it
is necessary for the GSE DAQ cards to simulate the rocket serial
interfaces for the three TM channels, including power interface and
command of the various event and status input channels, for laboratory
testing without the TM.

\section{Laboratory Characterization \& Equipment}
\label{S:labtesting}

The CIBER instrumentation is fully characterized in the lab before
flight using equipment described in this section.  Tests unique to
individual instruments are described in the CIBER Instrument Companion
papers.

\subsection{CIBER Optical Testing Configuration}
\label{sS:opticaltesting}

In flight the payload door opens directly to space and so does not use
a window; however, in the lab some means of introducing light into the
telescopes is necessary.  We use a customized window bulkhead for
optical testing in the lab with openings on which circular glass
windows are installed on o-ring seals, one for each instrument.  The
BK7 glass windows are measured to have transmittance $T > 0.85$ over
the range $400 < \lambda < 2000 \,$nm.

Calculations show that, due to the temperature difference between the
cryogenic insert and the outer envelope of the experiment section, the
radiative loading on the telescope apertures and front plate in lab
testing configuration is $\sim 20 \,$W.  Due to the finite thermal
conduction of the telescope lenses, under thermal loading the center
of a lens is significantly warmer than its edge so that the focus of
the instrument changes as a function of the loading.  Because in
flight configuration the loading on the first optic is $\sim 3 \,$W,
if left uncorrected the best focus position measured in the laboratory
would be significantly different than the best focus position in
flight.  To mitigate this, during optical tests an aluminum cold
window plate holding BK7 glass, $1/8''$ thick windows for each
instrument is installed.  If the cold window plate were to be
thermally sunk through the front plate and optical bench it would
cause the front of the optical bench to run at a higher temperature.
The cold window plate is therefore mechanically attached to the front
plate and thermally strapped to the radiation shield.  This
configuration allows the power from the warm skin to illuminate the
window plate and be sunk directly to the cryostat via the thermally
conductive radiation shields rather than through the instrument
apertures and optical bench.  In this testing configuration $\sim 5
\,$W of power falls on the telescope optics, which is close enough to
flight levels to allow good focus determination.  The cold window
design is successful as all four instruments were in good focus during
both of CIBER's flights.

\subsection{Collimators}
\label{sS:collimators}

The optical focus of the instruments is characterized using a
collimating telescope to mimic an astronomical point source.  Light
from a either a quartz-halogen lamp or monochromator (the latter is
discussed in Section \ref{sS:monochromator}) shines through a pinhole
placed at the collimating telescope’s focus, resulting in a collimated
beam.  Due to the different optical properties of the instruments, we
use two different collimating telescope systems.  For the
Spectrometers, the collimating telescope is a Parks $f/3.6$ (91 cm
focal length) Newtonian telescope\footnote{Manufactured by Parks
  Optical.}, while for the Imagers we use an $f/9.5$ (172 cm focal
length) off-axis Newtonian telescope\footnote{Fabricated in house
  using a primary mirror ground by DGM Optics.}.  An auto-collimating
microscope is used to set the focus of the collimators.  The best
focus position of the CIBER instruments is determined by moving the
collimating telescope's pinhole through its best focus position.
Using the thin lens approximation and given the focal lengths of the
instrument $f_{\mathrm{inst}}$ and collimating telescope
$f_{\mathrm{col}}$, the relation between shifts along the optical axis
at the position of the pinhole $\Delta l_{\mathrm{col}}$ and the
equivalent shift at the detector $\Delta l_{\mathrm{inst}}$ can be
calculated by:
\begin{equation} 
\Delta l_{\mathrm{inst}} = \left(
\frac{f_{\mathrm{inst}}}{f_{\mathrm{col}}} \right)^2 \Delta
l_{\mathrm{col}},
\end{equation}
where the smallest CIBER PSF is measured at $\Delta l_{\mathrm{col}}$.

Because each of the CIBER FPAs are mechanically fixed to their
respective instruments and it is not possible to determine the best
focus position of the focal planes a priori, the flight position of
the focal planes must be set experimentally.  This is done iteratively
by measuring the focus performance of the instrument thereby
determining the offset from best focus, warming the experiment,
shimming the focal plane in question to the revised focus position,
cooling back to LN$_{2}$ temperature, and remeasuring the focus
performance.  After several such cycles the design focus performance
of the instruments is achieved.

\vspace{0.5cm}

\subsection{Monochromator}
\label{sS:monochromator}

The spectral performance of each of the instruments must be
characterized in the lab prior to launch.  A commercial monochromator
system\footnote{Newport Corporation Oriel Monochromator and
  Spectrograph model number MS257 coupled to a Newport Corporation
  Apex 50 W quartz-tungsten-halogen lamp light source.} is used to
measure the spectral properties of the instruments.  The monochromator
utilizes two prisms with blaze wavelengths of $750$ and $1200 \,$nm
and rulings of $1200$ lines/mm and 400 lines/mm, respectively.
Together these gratings allow spectral coverage from $\sim 450 \,$nm
to $\sim 2500 \,$nm.  For a particular test the spectral band is
defined by selecting a fixed slit from a set of varying widths, and
the wavelength is selected by tuning the angle of incidence of the
light source on the grating.  The monochromator is calibrated at the
factory, and control software is provided by the manufacturer with the
calibration pre-loaded.  The calibration is checked periodically,
usually after shipping to or from the field, using a Ne lamp to
confirm the position of Ne emission lines matched with known spectra.
Absolute wavelength accuracies $< 1 \,$nm and repeatabilities $< \pm
0.5 \,$nm are readily achieved with this system.  The output port of
the monochromator can be coupled to the pinhole of a collimating
telescope if a monochromatic, quasi-point point source is desired, or
coupled to the optical fiber input of an integrating sphere if a
monochromatic field-filling source is necessary: both configurations
are used for CIBER testing.  The brightness of the monochromator
output can be attenuated using any combination of filters and
geometric techniques as required.

\subsection{Integrating Sphere \& Vacuum Box}
\label{sS:labsphere}

\begin{figure}[htb]
\begin{center}
\epsfig{file=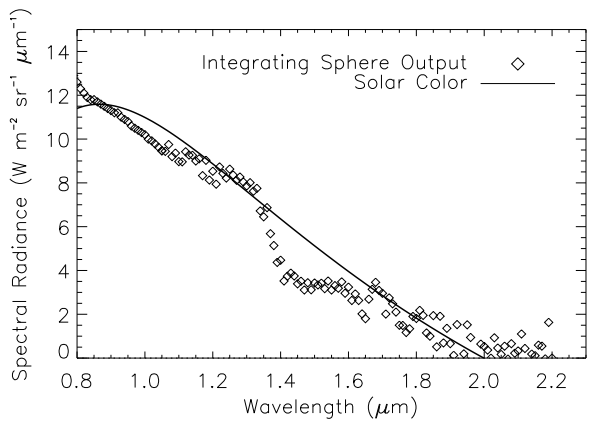,width=0.4\textwidth}
\caption{Output unattenuated spectral radiance of the integrating
  sphere used as a beam filling source for calibrating the CIBER
  instruments.  The spectrum is designed to be close to solar in the
  range $800 < \lambda < 2200 \,$nm to mimic the expected color of the
  Zodiacal light which increases the fidelity of the flat field
  measurements taken with this system.  A scaled solar spectrum is
  shown for reference.  The decrement near $1.4 \, \mu$m is likely due
  to absorption from a water line in the atmosphere.  The latter
  effect is mitigated by performing measurements with the sphere in an
  evacuated chamber.}
\label{fig:spherespec}
\end{center}
\end{figure}

An integrating sphere is used to measure the CIBER instrument response
to diffuse sources which illuminate the entire $A \Omega$ of the
telescopes such as the CIRB or Zodiacal light.  We employ a commercial
integrating sphere system\footnote{Manufactured by Labsphere, Inc.}
with a $20 \inchsign$ diameter and an $8 \inchsign$ exit port which
can be stopped down with port reducers if desired.  The interior of
the sphere is coated with a barium sulfate paint and the sphere
assembly is attached to a frame mount for ease of positioning.  The
light source for the sphere is a quartz tungsten halogen
lamp\footnote{Manufactured by Schott North America, Inc.} in a housing
with a customized filter designed to produce an approximately solar
spectrum at the sphere's output.  Figure \ref{fig:spherespec} shows
the measured spectrum of the integrating sphere system compared to a
solar spectral shape.

The output of the light source is coupled to the sphere using a 6'
long optical fiber bundle.  By adjusting the diameter of a pinhole at
the light source, the absolute radiance presented by the large sphere
can be modified without changing the output spectrum.  In addition, a
smaller $4\inchsign$ diameter injection integrating sphere is coupled
to the larger sphere and illuminates the pinhole with its output port,
attenuating the input radiance and improving the larger sphere's
output uniformity.  A monitor detector\footnote{This detector is
  either an Si or InGaAs photodiode, depending on the instrument under
  test.} is attached to the injection sphere and is readout using a
commercial pre-amplifier and display system\footnote{Manufactured by
  Labsphere Inc.}.  The integrating sphere system is measured to have
output radiance uniform to $< 2$\% at angles up to $15^{\circ}$ from
the center line of the output port and its spectral radiance can be
adjusted continuously from $\sim 10$ to $\sim 10^{-8}
$W/m$^2$/sr/$\mu$m using the injection sphere and a selection of
pinhole attenuators.

\begin{figure}[htb]
\begin{center}
\epsfig{file=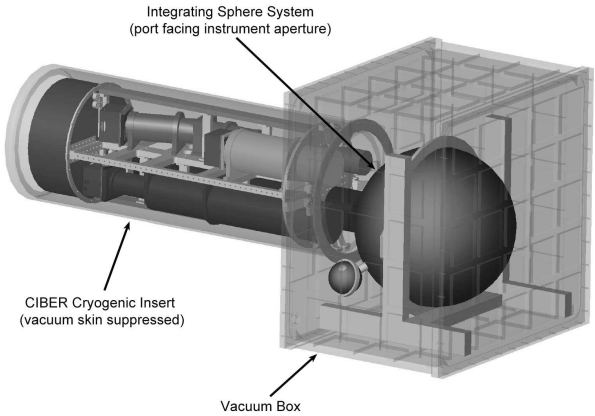,width=0.45\textwidth}
\caption{A schematic view of the integrating sphere inside the vacuum
  box coupled to CIBER.  The outer skins of the vacuum sections have
  been made transparent.  The vacuum box is constructed from Al-2024,
  chosen for its tensile strength and light weight.  To reduce the
  weight of the box further, a grid of $\sim 4 \inchsign$ square, $0.5
  \inchsign$ deep pockets are milled out of each of the walls, which
  reduces the total weight of the box by $50$\%.  The sphere is
  positioned such that a particular CIBER instrument is viewing its
  output port and measurements are taken with CIBER operating at $77
  \,$K.  Electrical monitoring signals and light source input are
  passed through hermetic connections on the front plate of the box.}
\label{fig:vacuumbox}
\end{center}
\end{figure}

Laboratory measurements of the instruments' flat field responsivity
are complicated both by the presence of the warm vacuum windows and by
absorption by water in the atmosphere.  The former is a potentially
complex spatial and spectral correction while the latter can attenuate
the near-IR spectrum in broad regions around $1.4$ and $1.9 \,\mu$m.
To remove these effects it is necessary to perform the flat field
measurements in vacuum.  To this end, a $36 \times 36 \times 24$ cubic
inch vacuum sphere was made.  It is constructed of $0.75 \inchsign$
thick Al-2024 plates\footnote{The box was fabricated at the Korea
  Basic Science Institute in Daejeon, Korea.} and is shown
schematically in Figure \ref{fig:vacuumbox}.  The interior of the
vacuum box is painted with Aeroglaze Z306\footnote{Manufactured by the
  Lord Corporation.} to absorb scattered light.  The box can be pumped
out independently of the experiment section if required, and a
hermetic electrical interface allows electrical signals to be passed
to the interior of the box.  In addition, a separate hermetic optical
fiber bundle is used to couple an external light source into the
integrating sphere.  This allows real time control of the input light
entering the sphere from outside of the vacuum box.

\subsection{Calibration}
\label{sS:calibration}

Both Spectrometers require a calibration of their absolute
responsivity to extended emission.  We opt to measure a calibrated
extended source in the laboratory to calibrate the CIBER instruments
using instruments developed at the National Institute of Standards and
Technology.

\begin{figure*}[htb]
\begin{center}
\epsfig{file=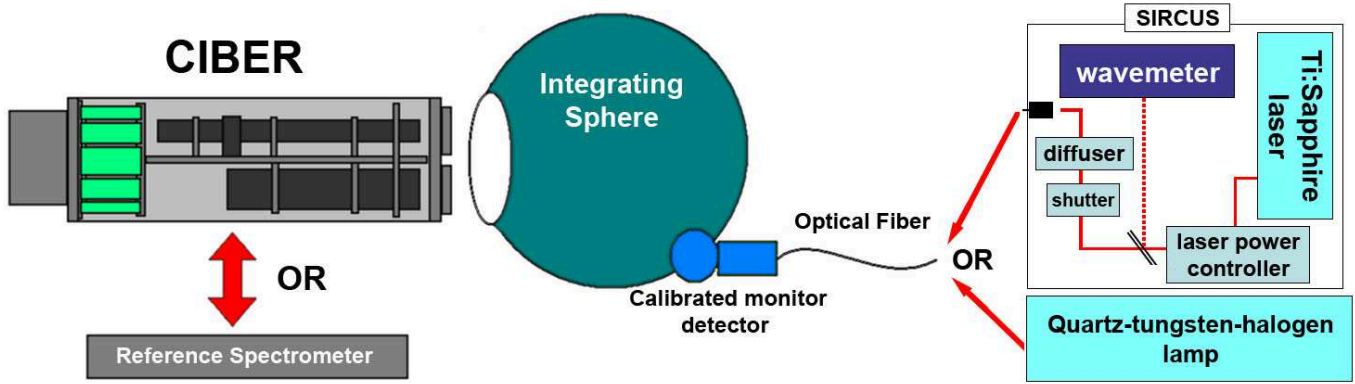,width=1.0\textwidth}
\caption{Schematic diagram of the experimental setup for the CIBER
  instrument absolute calibration using the either SIRCUS facility
  (described in \citealt{Brown2006} and shown here in broad schematic
  view) or a commercial quartz-tungsten-halogen white light source.
  The source is fiber coupled into a 4' diameter integrating sphere
  with a $18 \inchsign$ port; the output of the sphere is viewed by
  either the instrument under test or a reference irradiance detector.
  Specifics of the calibration of each instrument are presented in the
  CIBER instrument companion papers.}
\label{cal_config}
\end{center}
\end{figure*}

A schematic diagram of the CIBER calibration apparatus is shown in
Figure~\ref{cal_config}.  Depending on the desired calibration
product, two measurements using different light sources are available:
the first uses the Spectral Irradiance and Radiance Calibrations using
Uniform Sources (SIRCUS) facility \citep{Brown2006} while the second a
standard quartz-tungsten-halogen lamp.  The SIRCUS facility utilizes a
Nd:Vanadate-pumped Ti:sapphire tunable laser for monochromatic
measurements.  In both cases, the light source is coupled to a $4'$
diameter integrating sphere whose port is viewed by CIBER through the
vacuum window bulkhead.  The transfer standard detectors are different
for the two sets of measurements: a NIST-calibrated Si radiance meter
is used for the laser measurements, and the NIST Remote Sensing
Laboratory's Analytical Spectral Devices (ASD) FieldSpec3
spectroradiometer is used for the white light measurements.  These
measurements are used to accurately calibrate both the absolute and
relative (i.e.~pixel-to-pixel) responsivity; specific details can be
found in the CIBER Instrument Companion papers.

\section{CIBER in Flight}
\label{S:flight}

The CIBER experiment section is designed to launch on the Black Brant
IX vehicle (\citealt{SRPH2005}, \citealt{Krause2005}).  From aft to
forward, the CIBER payload stack consists of the Terrier booster, the
Black Brant sustainer, the ignitor for the motors, a spring loaded
separation section, the experiment section as described in this paper,
the star tracker housing, the telemetry section, the active guidance
section, the celestial attitude control system, and finally the
recovery system in the nose cone.  Each section is housed in its own
modular section, and the payload is a uniform $17.26$ inch diameter
along its length.  Together, these systems provide launch and
separation capability, active pointing while in flight, real time
telemetry, and parachute recovery.  Except for the telemetry
(discussed in Section \ref{sS:TMGSE}) and star tracker sections, which
are customized for the experiment, the sections are all standardized
and can be swapped with spares as required.  At launch, the fully
integrated CIBER payload is $5.82 \,$m long and weighs $450
\,$kg\footnote{The weight of the payload may vary by $\pm 10 \,$kg
  depending on the balancing weights required in a given flight.}.

The star tracker section uses a University of Wisconsin developed model
ST-5000 side-looking star tracker \citep{Percival2008}.  The CIBER
payload is the first deployment of this system, whose advantage is
that it need not be co-mounted at the instrument aperture (and thus
take aperture space from the science instruments).  The side-looking
star tracker is aligned with the instruments using a secondary,
temporarily installed ST-5000 system which is co-mounted to the aft
bulkhead.  This temporary tracker is optically aligned to the
instruments using a collimator, and the resulting pointing solution is
transferred to the side-looking tracker by measuring star positions in
both trackers simultaneously.  Stack-up tolerance, and heating and
vibrational effects at launch can lead to error in the absolute
pointing model of $\lesssim 15'$.

\begin{figure}[htb]
\begin{center}
\vspace{0.5cm}
\epsfig{file=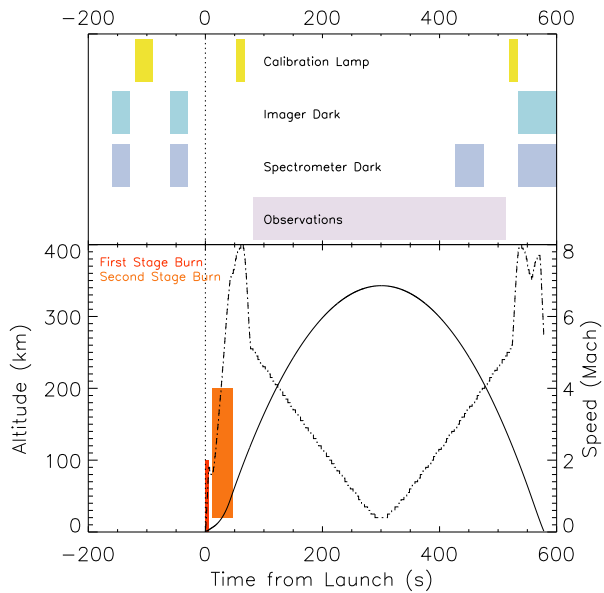,width=0.4\textwidth}
\caption{The typical flight profile for CIBER on a Black Brant IX
  sounding rocket.  Prior to flight, the cold shutters and calibration
  lamps are operated to provide a baseline measurement of the
  instrument performance.  At $t=0$, the first stage burns for $\sim 6
  \,$s and the second for $\sim 30 \,$s.  The door opens and
  astrophysical observations begin at about $130 \,$km; $\sim 420 \,$s
  of observation time can be split between as many as 10 astronomical
  targets depending on the scientific drivers of the flight.  As
  knowledge of the absolute level of their detectors' current is key,
  towards the end of the observation period, the shutters are closed
  on the Spectrometers only.  Finally, the door closes at $\sim 520
  \,$s, the calibration lamp is flashed and the cold shutters are
  closed for re-entry.}
\label{fig:flightprofile}
\end{center}
\end{figure}

The CIBER flight profile is summarized in Figure
\ref{fig:flightprofile}.  At launch, the rocket burns its Mark 70
Terrier first stage; the burn lasts for approximately $6 \,$s and then
the motor is drag-separated from the second stage.  Guide fins at the
base of the Terrier motor are canted by a small angle to give the
rocket an $\sim 4 \,$Hz roll rate, stabilizing the rocket along its
long axis.  The second stage sustainer, a Black Brant mod 2, burns for
approximately $30 \,$s.  Immediately following the second burn, most
of the angular momentum acquired during the Terrier burn is shed by
utilizing two cables with weights at their ends to de-spin the
payload.  These are released and, when the cables reach their full
extent, are cut to cast off angular momentum, following which the
payload is spinning at $\sim 0.5\, $Hz.  Next, the spent sustainer is
pushed away from the payload by the high velocity spring section
attached to the top of the motor housing.  This ensures that the motor
and payload are separated by a large distance when data acquisition
commences, as previous experiments have become contaminated with
residual motor exhaust at altitude (see \citealt{Matsuura1994},
\citealt{Yost2000} for a discussion of this problem).  The experiment
is then fully despun to $0 \,$Hz by pressurized gas controlled by the
ACS, the shutter door opens, and scientific observations begin.  The
payload pointing slews from field to field by the ACS-controlled
pressurized gas jets, achieving a maximum slew rate of $\sim
40^{\circ}$/s.  Fine pointing is governed by gyroscopes.  At the end
of the flight prior to re-entry, the experiment door is closed and gas
jets are used to spin the payload back up to $\sim 1 \,$Hz to reduce
the frictional heating on any single area of the payload.  The
parachute deploys at an altitude of $16,000 \,$ft and the payload
impacts the ground at about $30 \,$ft/s.  The payload is tracked along
its entire trajectory and a recovery team is deployed the following
morning to retrieve the payload.

The choice of our science fields was the same in both of the first two
flights (and is discussed in detail in \citealt{Tsumura2010}), but is
in principle entirely flexible based on requirements of the science
team's choosing.  Fields covered by large area optical and near IR
surveys are chosen for the extragalactic fluctuation science while
fields at a variety of ecliptic latitudes are chosen for the Zodiacal
light characterization.  This leads to a flight profile for which $\sim
25\,$\% of the observation time is spent on Zodiacal light science and
the remainder on deep cosmological fields.

A set of four criteria define CIBER's flight windows based on
astronomical considerations.  Firstly, every field must have ZA$ <
50^{\circ}$ to reduce the possibility for Earth-shine contamination.
Secondly, the zenith angle (ZA) of the Sun must be $>113.6^{\circ}$ so
that the Sun is $5^{\circ}$ below the depressed horizon.  Thirdly, the
ZA of the moon must be $>108.6^{\circ}$ so that the moon is below the
depressed horizon.  Finally, to ensure that the CIBER data matches the
conditions under which the DIRBE data was taken, the solar elongation
(SE) to each of the fields must be in the range $64^{\circ} <
\mathrm{SE} < 124^{\circ}$.  Together, these conditions tend to
restrict CIBER flight windows to a few $\sim 10$ day periods clustered
in the middle of winter and early summer months from North American
launch sites.

CIBER has undertaken two successful flights using a Terrier-Black
Brant IX rocket launched from White Sands Missile Range, first on 25
February, 2009 and then on 10 July, 2010.  In both cases the payload
had a remarkably similar flight, achieving an apogee of $\sim 330
\,$km and providing $>420 \,$s of astronomical data.  In both flights
the calibration lamps and cold shutters were successfully operated
during the ascent and descent phase to monitor the performance of the
detectors under flight conditions.  The payload pointing stability was
$< 3''$ over $\sim 30 \,$s integrations in a given field.  In both
cases, the instrument was successfully recovered for future flights.
\citet{Tsumura2010} presents initial results from the first flight;
analysis of the second flight data is currently underway and
scientific results are expected soon.  As the data quality from the
second flight appears excellent, no major hardware modifications are
expected for future CIBER-1 flights.  Beyond that, CIBER-2, a
completely redesigned follow-on payload to CIBER-1 optimized for
reionization fluctuation science, is in its design phase and is
expected to fly in 2014.


\section*{Acknowledgments} 

This work was supported by NASA APRA research grants NNX07AI54G,
NNG05WC18G, NNX07AG43G, NNX07AJ24G, and NNX10AE12G.  Initial support
was provided by an award to J.B.~from the Jet Propulsion Laboratory's
Director's Research and Development Fund.  Japanese participation in
CIBER was supported by KAKENHI (20$\cdot$34, 18204018, 19540250,
21340047 and 21111004) from Japan Society for the Promotion of Science
(JSPS) and the Ministry of Education, Culture, Sports, Science and
Technology (MEXT).  Korean participation in CIBER was supported by the
Pioneer Project from Korea Astronomy and Space science Institute
(KASI).

We would like to acknowledge the dedicated efforts of the sounding
rocket staff at the NASA Wallops Flight Facility and the White Sands
Missile Range.  We also acknowledge the work of the Genesia
Corporation for technical support of the CIBER optics, and M.C.~Runyan
and an anonymous referee for helpful comments on this manuscript.
M.Z.~acknowledges support from a NASA Postdoctoral Fellowship,
A.C.~acknowledges support from an NSF CAREER award, B.K.~acknowledges
support from a UCSD Hellman Faculty Fellowship, and K.T.~acknowledges
support from the JSPS Research Fellowship for Young Scientists.


\bibliography{ms}

\end{document}